\documentclass[preprint,showpacs,titlepage,aps,prd,tightenlines,amsmath,byrevtex,nofootinbib]{revtex4}

\usepackage{graphicx}
\usepackage{epsfig}\parskip 5pt plus 1pt
\usepackage{color}

\definecolor{MyGrey}{rgb}{0,0,0} 
\definecolor{MyDarkBlue}{rgb}{0.3,0.3,0.9} 
\definecolor{MyLightBlue}{rgb}{0.22,0.51,0.9}

\usepackage[colorlinks=true,linkcolor=MyDarkBlue,citecolor=MyDarkBlue,urlcolor=MyLightBlue,bookmarksnumbered=true,bookmarksopen]{hyperref}
\usepackage{breakurl} 

\def\lsim{\raise0.3ex\hbox{$\;<$\kern-0.75em\raise-1.1ex
\hbox{$\sim\;$}}}
\def\gsim{\raise0.3ex\hbox{$\;>$\kern-0.75em\raise-1.1ex
\hbox{$\sim\;$}}}

\newcommand{\sth}{\sin^2\theta_{23}}

\newcommand{\be}{\begin{equation}}
\newcommand{\ee}{\end{equation}}
\newcommand{\bea}{\begin{eqnarray}}
\newcommand{\eea}{\end{eqnarray}}

\newcommand{\ie}{{\it i.e.}}
\newcommand{\eg}{{\it e.g.}}

\begin{document}
\vspace*{-1cm}
\title{Interplay between Appearance and Disappearance Channels for Precision Measurements of $\theta_{23}$ and $\delta$}


\author{Pilar~Coloma$^{1}$}
\email{pcoloma@vt.edu}
\author{Hisakazu~Minakata$^{2}$}
\email{hisakazu.minakata@gmail.com}
\author{Stephen~J.~Parke$^{3}$}
\email{parke@fnal.gov} 
\affiliation{
$^1$Center for Neutrino Physics, Virginia Tech, Blacksburg, VA 24061, USA \\
$^2$Instituto de F\'{\i}sica, Universidade de S\~ao Paulo, C.\ P.\ 66.318, 05315-970 S\~ao Paulo, Brazil \\ 
$^3$Theoretical Physics Department, Fermi National Accelerator Laboratory, P.\ O.\ Box 500, Batavia, IL 60510, USA \\ }
\preprint{FERMILAB-PUB-14-115-T}
\preprint{NORDITA-2014-54}

\date{\today}

\vglue 1.6cm

\begin{abstract}

We discuss how the CP violating phase $\delta$ and the mixing angle $\theta_{23}$ can be measured precisely in an environment where there are strong correlations between them. This is achieved by paying special attention to the mutual roles and the interplay between the appearance and the disappearance channels in long-baseline neutrino oscillation experiments. We analyze and clarify the general structure of the $\theta_{23} - \theta_{13} - \delta$ degeneracy for both the appearance and disappearance channels in a more complete fashion than what has previously been discussed in the literature. A full understanding of this degeneracy is of vital importance if $\theta_{23}$ is close to maximal mixing. The relative importance between the appearance and disappearance channels depends upon the particular setup and how close to maximal mixing Nature has chosen the value for $\theta_{23}$. For facilities that operate with a narrow band beam or a wide band beam centered on the first oscillation extremum, the contribution of the disappearance channel depends critically on the systematic uncertainties assumed for this channel. Whereas for facilities that operate at energies above the first oscillation extremum or at the second oscillation extremum the appearance channels dominate.  On the other hand, for $\delta$ we find that the disappearance channel usually improves the sensitivity, modestly for facilities around the first oscillation extremum and more significantly for facilities operating at an energy above the first oscillation extremum, especially near $\delta \sim \pm \pi/2$.

\end{abstract}


\maketitle

\section{Introduction}

The three flavor mixing angles in the lepton sector are all measured now and the next step is to measure the CP violating phase $\delta$~\cite{Nunokawa:2007qh}. It would be the last step, aside from determination of the neutrino mass hierarchy, to complete our understanding of lepton mixing in the standard three generation scheme. Lepton CP violation is one of the indispensable ingredients for leptogenesis~\cite{Fukugita:1986hr} which could explain baryon number asymmetry in the universe. 

At the same time neutrino physics is entering the precision era. Precision will help successful model building in the leptonic sector, which eventually should lead to the resolution of the so-called flavor puzzle. 
Fortunately, $\theta_{13}$ will be soon determined accurately by the Daya Bay and the other reactor experiments~\cite{An:2013zwz,Abe:2012tg,fortheRENO:2013lwa}, which is free from the uncertainties on $\theta_{23}$ and $\delta$~\cite{Minakata:2002jv}. 
Given the high accuracy of $\theta_{12}$ measurement by the solar~\cite{Aharmim:2011vm} (see also e.g., \cite{Robertson:2006pk} for a review on results from solar oscillation experiments) and the KamLAND~\cite{Gando:2013nba} experiments, which may be even more improved at the JUNO~\cite{Li:2013zyd} or RENO-50~\cite{ShinA:2014iea} experiments, $\theta_{23}$ will be the least precisely known mixing angle. 
Then, the uncertainty of $\theta_{23}$ could be one of the dominant sources of uncertainty for the measurement of $\delta$, in addition to statistical and systematic ones. 

Up to now, it is generally assumed that $\theta_{23}$ will mainly be determined through $\nu_{\mu}$ disappearance measurements, and $\delta$ is to be measured by $\nu_e$ and $\bar{\nu}_e$ appearance measurements, possibly simultaneously with $\theta_{13}$, or with a given measured value of $\theta_{13}$ by reactor experiments. However, it turns out that the problem of determining $\theta_{23}$ and $\delta$ simultaneously is not that simple. 

If $\theta_{23}$ is close to maximal mixing, \ie\ $\sin^2 2\theta_{23} \gsim 0.96$ (a value to which the experimental results seem to be converging), the determination of $\sin^2 \theta_{23}$ will be difficult because the two allowed regions for $\theta_{23}$ (the true solution and one clone) merge together~\cite{Minakata:2004pg}. As a result, the final allowed region for $\sin^2 \theta_{23}$ will span both the first and second octants of $\theta_{23}$. 
It was shown that the $\nu_e$ and $\bar{\nu}_e$ appearance measurements by themselves, no matter how accurate, produce a continuous ``tusk shaped'' degeneracy line, parameterized by the value of $\delta$, in $\sin^2 \theta_{13} - \sin^2 \theta_{23}$ space~\cite{Minakata:2002jv}. Though the three-dimensional $\theta_{23} - \theta_{13} - \delta$ parameter space is squeezed by the reactor measurement of $\theta_{13}$ (yet with finite resolution), we still have to deal with the problem of determining $\theta_{23}$ and $\delta$ simultaneously~\cite{Minakata:2013eoa}. We will see that it suffers from a parameter degeneracy involving $\theta_{23}$, $\theta_{13}$, and $\delta$. 

We utilize the following four experimental setups to illuminate the characteristic features of this degeneracy:
\noindent
(1) T2HK~\cite{Abe:2011ts} for a representative case of setups whose neutrino spectrum is peaked near the first vacuum oscillation maximum (VOM), such that 
$\Delta_{31} \equiv  (m^2_{3} - m^2_1) L/4 E = \pm \pi/2$, where $L$ is the distance to the detector and $E$ is the neutrino energy,
(2) LBNE~\cite{Adams:2013qkq} for a representative case of setups with wide-band neutrino beams around $ \vert \Delta_{31} \vert \sim   (2\pm1) \pi/4$,
(3) Neutrino Factory (NF)~\cite{Agarwalla:2010hk,idsnfweb} 
for a representative case of setups at higher energies than VOM, $ \vert \Delta_{31} \vert  \sim  \pi/4$, and at long baseline with sizable matter effect, and 
(4) ESS$\nu$SB~\cite{Baussan:2012cw,Baussan:2013zcy} for a representative case of setups with neutrino spectrum peaked near the second VOM, $ \vert \Delta_{31} \vert =  3 \pi/2$.

The paper is structured as follows. First, we aim at illuminating the structure of the parameter degeneracy in the three-dimensional $\theta_{23} - \theta_{13} - \delta$ space, which will be denoted as the general $\theta_{23} - \theta_{13} - \delta$ degeneracy. Despite that the intrinsic $\theta_{13}$-$\delta$ degeneracy~\cite{BurguetCastell:2001ez} 
(and to less extent the intrinsic $\theta_{23}$-$\delta$ one \cite{Minakata:2013eoa}) 
multiplied with the discrete $\theta_{23}$ octant degeneracy~\cite{Fogli:1996pv} (see also~\cite{Barger:2001yr}) has been discussed extensively in the literature,  
to our knowledge, its full structure has never been addressed in a complete fashion. A detailed discussion of this degeneracy will be presented in Sec.~\ref{sec:G-degeneracy}. Then, in the rest of the paper, we study how well can $\theta_{23}$ and $\delta$ be measured at future neutrino oscillation facilities, focusing in particular on the relative importance of $\nu_{\mu}$ disappearance vs. $\nu_\mu \rightarrow \nu_e$ ($\bar{\nu}_\mu \rightarrow \bar{\nu}_e$) appearance measurements\footnote{
Here, and in the rest of this work, we denote the appearance channels for the super beam experiments with the implicit understanding that for the Neutrino Factory or Beta Beam experiments the appearance channels are $\nu_e \rightarrow \nu_\mu$ and its CP conjugate.} 
%
for a precise determination of $\sin^2 \theta_{23}$ and $\delta$. This will be discussed in Secs.~\ref{sec:relative} and \ref{sec:delta}, respectively. Finally, we summarize our results and present our conclusions in Sec.~\ref{sec:conclusions}. 

\section{General $\theta_{23} - \theta_{13} - \delta$ Degeneracy of the Appearance and Disappearance Channels}
\label{sec:G-degeneracy}

Here, we discuss the general structure of degeneracy involving $\theta_{23}$, $\theta_{13}$, and $\delta$ which is  encountered in the measurement of these parameters. Our aim in this section is to illuminate the nature of this parameter degeneracy but not to go deeply into discussing how it can be resolved. However, we do expect that our discussion will be useful to formulate the resolution of this degeneracy. While our discussions in this section are meant to be pedagogical in nature, many of the features of this general $\theta_{23} - \theta_{13} - \delta$ degeneracy are entirely new.

For the sake of simplicity, we will assume throughout this paper that the neutrino mass hierarchy is known to be the normal hierarchy. In the case of unknown mass hierarchy, the number of allowed solutions would be doubled since the clone solutions would also appear for the wrong mass hierarchy~\cite{Minakata:2001qm}. The extension can be done in a straightforward manner. Finally, the inclusion of matter effects complicates the discussion without adding too much to the understanding. Therefore we will turn them off in the rest of this section.

\subsection{Observables and Overview}
\label{sec:observables}

In this paper, we consider the following four observables in discussing the determination of $\theta_{23}$, $\theta_{13}$, and $\delta$: 
\begin{enumerate}

\item 
$P_{\mu e} (\theta_{23}, \theta_{13}, \delta) $ and $\bar{P}_{\mu e} (\theta_{23}, \theta_{13}, \delta)$:  the appearance oscillation probabilities\footnote{Here only the variables which have important effect in our discussion are shown as arguments of the oscillation probabilities. Explicit expressions for the oscillation probabilities will be given below.
} for $\nu_{\mu} \rightarrow \nu_{e}$ and $\bar{\nu}_{\mu} \rightarrow \bar{\nu}_{e}$ respectively. For these probabilities there is a continuous degeneracy in the three variables  $\theta_{23}$, $\theta_{13}$ and $\delta$. We will refer to this degeneracy as the  ``$\theta_{23} - \theta_{13} - \delta$ appearance degeneracy.''

\item 
$P_{\mu\mu} (\theta_{23}, \theta_{13})$: the disappearance oscillation probability for $\nu_{\mu} \rightarrow \nu_{\mu}$. For this probability there is a continuous degeneracy in the two variables  $\theta_{23}$ and $\theta_{13}$. We will refer to this degeneracy as the ``$\theta_{23} - \theta_{13}$  disappearance degeneracy.''

\item 
$\bar{P}_{ee} (\theta_{13})$: the disappearance oscillation probability for $\bar{\nu}_{e} \rightarrow \bar{\nu}_{e}$. There is no degeneracy in this channel since $\cos^2 \theta_{13}$ is not small and therefore $\theta_{13}$ is determined unambiguously. (See discussion after Eq.~\ref{Pee}.)

\end{enumerate}

In this section we restrict ourselves to the analytic treatment of the degeneracy assuming measurements of the above observables for a fixed neutrino energy $E$. Since there are four equations for the three variables, the system is, in general, over constrained and in principle there is no degeneracy if each measurement is precise enough except at possible isolated values of the neutrino energy.  However, degeneracies may appear if the measurements are not accurate enough.

Let us start by discussing what has been addressed in the literature up to now regarding the degeneracies associated with  $\theta_{23} - \theta_{13} - \delta$ .

\begin{itemize}

\item[(a)] If a $\nu_{\mu}$ disappearance measurement of $\sin^2 2 \theta_{23}$ is sufficiently accurate to determine $\theta_{23}$ (up to its octant), then a set of $\nu_e$ and $\bar{\nu}_e$ appearance measurements would give two allowed solutions for ($\theta_{13}$, $\delta$): the true solution and a degenerate one, which has been referred to as ``$\theta_{13}$ intrinsic degeneracy''~\cite{BurguetCastell:2001ez}. Moreover, one would get two solutions for each value of $\theta_{23}$; thus, this degeneracy is fourfold (eightfold if we consider that the sign of $\Delta m^2_{31}$ is unknown).

\item[(b)] If the accuracy in determining $\sin^2 \theta_{13}$ overwhelms that of $\sin^2 \theta_{23}$, which is more or less the case after the reactor measurement of $\theta_{13}$, a set of $\nu_e$ and $\bar{\nu}_e$ appearance measurements would give two allowed solutions for ($\theta_{23}$, $\delta$): the true solution and a degenerate one, which has been referred to as the ``$\theta_{23}$ intrinsic degeneracy''~\cite{Minakata:2013eoa}.  In this case, the degeneracy is twofold excluding the ambiguity of the mass hierarchy.

\end{itemize}
\noindent 
In the first case, (a) above, the resultant fourfold degeneracy has been described as a direct product of the $\theta_{13}$-intrinsic and the $\theta_{23}$ octant degeneracies. Whereas the second case (b) is a $\theta_{23}$-intrinsic degeneracy which could in principle be resolved by an accurate determination of $\sin^2 \theta_{23}$ from a $\nu_\mu$ disappearance experiment. However, if $\theta_{23}$ is near maximal mixing ($\sim \frac{\pi}{4}$) this is challenging due to the Jacobian involved in translating the measured variable $\sin^2 2\theta$ to $\sin^2 \theta$.  

The general $\theta_{23} - \theta_{13} - \delta$ degeneracy we discuss here is best considered to be made up of two separate degeneracies: one associated with the appearance channels, the $\theta_{23} - \theta_{13} - \delta$ appearance degeneracy; and the second associated with the $\nu_\mu$ disappearance channel, the $\theta_{23} - \theta_{13}$ disappearance degeneracy. Both these degeneracies are continuous in the associated variables and will be illuminated in more detail in the following subsections.

\subsection{ Appearance Channels and the $\theta_{23} - \theta_{13} - \delta$ Appearance Degeneracy}
\label{sec:Appearance}

We start by describing the $\nu_{e}$ and $\bar{\nu}_{e}$ appearance measurements to understand the structure of $\theta_{23} - \theta_{13} - 
\delta$ appearance degeneracy. We will show that both the $\theta_{23}$ and the $\theta_{13}$ intrinsic degeneracies can be identified as particular projections of this appearance degeneracy.

The $\nu_{e}$ and $\bar{\nu}_{e}$ appearance oscillation probabilities can be written as
\begin{eqnarray}
P(\nu_{\mu} \to \nu_e) &=& 
\left( s_{23} \sin 2\theta_{13} \right)^2 
A_{\oplus}^2 
+ 2 \epsilon  \left( s_{23} \sin 2\theta_{13} \right) \left( c_{23} c_{13} \right) 
A_{\oplus} A_{\odot} \cos \left(  \delta+ \Delta_{31} \right) 
\nonumber \\
& &   +
~\epsilon^2  (c_{23} c_{13})^2 A_{\odot}^2, 
\nonumber \\
P(\bar{\nu}_{\mu} \to \bar{\nu}_e) &=& 
\left( s_{23} \sin 2\theta_{13} \right)^2
\bar{A}_{\oplus}^2 + 2 \epsilon  \left( s_{23} \sin 2\theta_{13} \right) \left( c_{23} c_{13} \right) 
\bar{A}_{\oplus} A_{\odot} \cos \left(  \delta- \Delta_{31} \right) 
\nonumber \\
& &  +
~\epsilon^2 (c_{23} c_{13})^2 A_{\odot}^2, 
\label{Pemu-matter}
\end{eqnarray}
where $\Delta_{ij} \equiv \frac{ \Delta m^2_{ij} L}{4E}$, $\epsilon \equiv \frac{ \Delta m^2_{21} } { \Delta m^2_{31} } \simeq 0.03$.  
The $A$ functions in (\ref{Pemu-matter}) are defined\footnote{Our definition of the $A$ functions differs from that of reference \cite{Minakata:2013eoa}.} as 
\begin{eqnarray}
A_{\oplus} &\equiv& 
 \left( \frac{ \Delta m^2_{31} }{ \Delta m^2_{31} - a } \right) 
\sin \left[ \frac{ ( \Delta m^2_{31} - a) L}{4E} \right], 
\nonumber \\
\bar{A}_{\oplus} &\equiv& 
\left( \frac{ \Delta m^2_{31} }{ \Delta m^2_{31} + a } \right) 
\sin \left[ \frac{ ( \Delta m^2_{31} + a) L}{4E} \right], 
\nonumber \\
A_{\odot} &\equiv& \sin 2\theta_{12}
\left( \frac{ \Delta m^2_{31} } { a } \right) 
\sin \left(\frac{ a L}{4E}  \right) = \bar{A}_{\odot}.
\label{A-def}
\end{eqnarray}
Here, $a = 2\sqrt{2} G_F N_e E$, where $G_F$ is the Fermi constant, $N_e$ is the electron density in matter and $E$ is the neutrino energy.

\begin{figure}[b]
\begin{center}
\vspace{-3mm}
\includegraphics[width=0.415\textwidth]{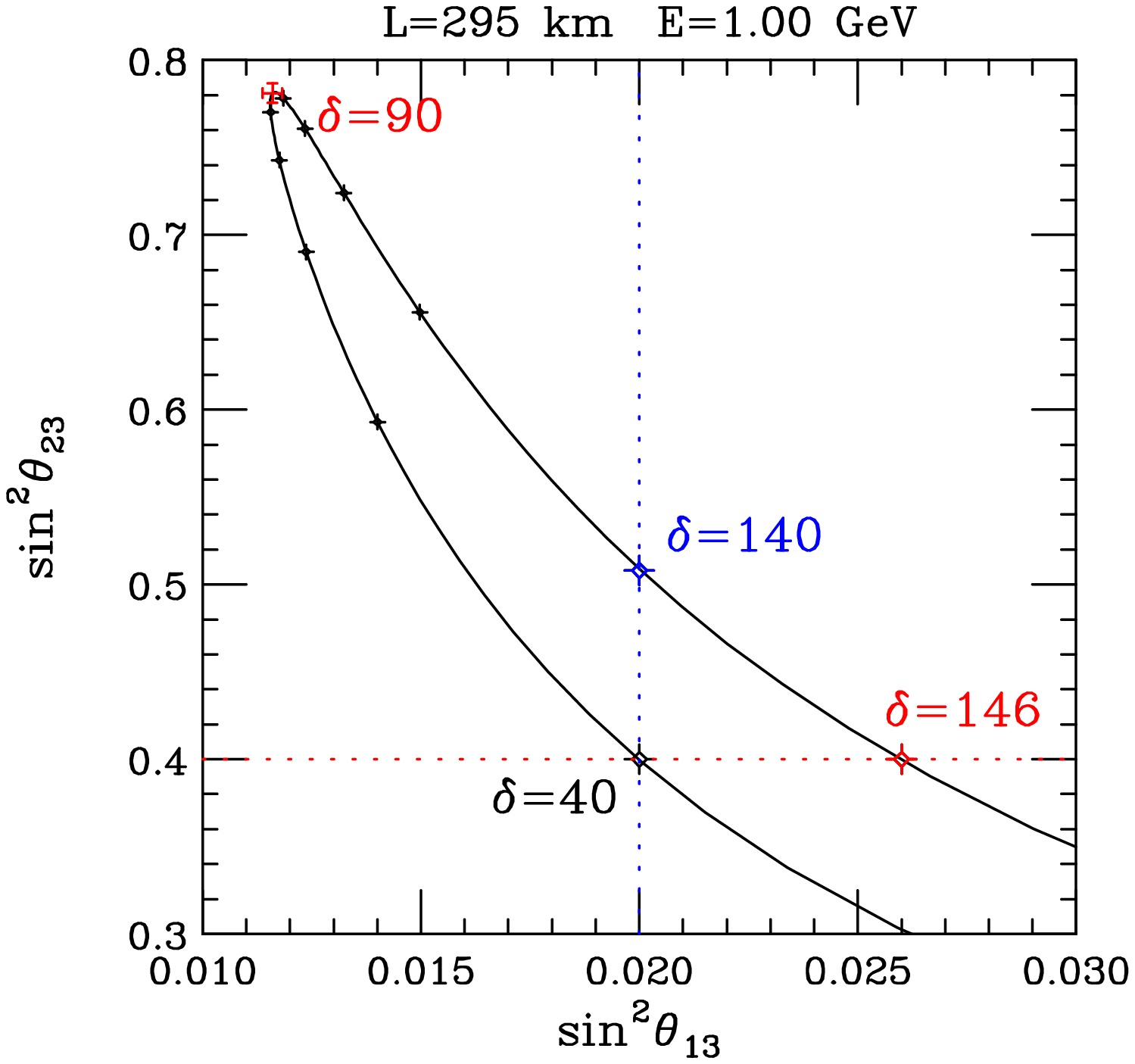}
\includegraphics[width=0.40\textwidth]{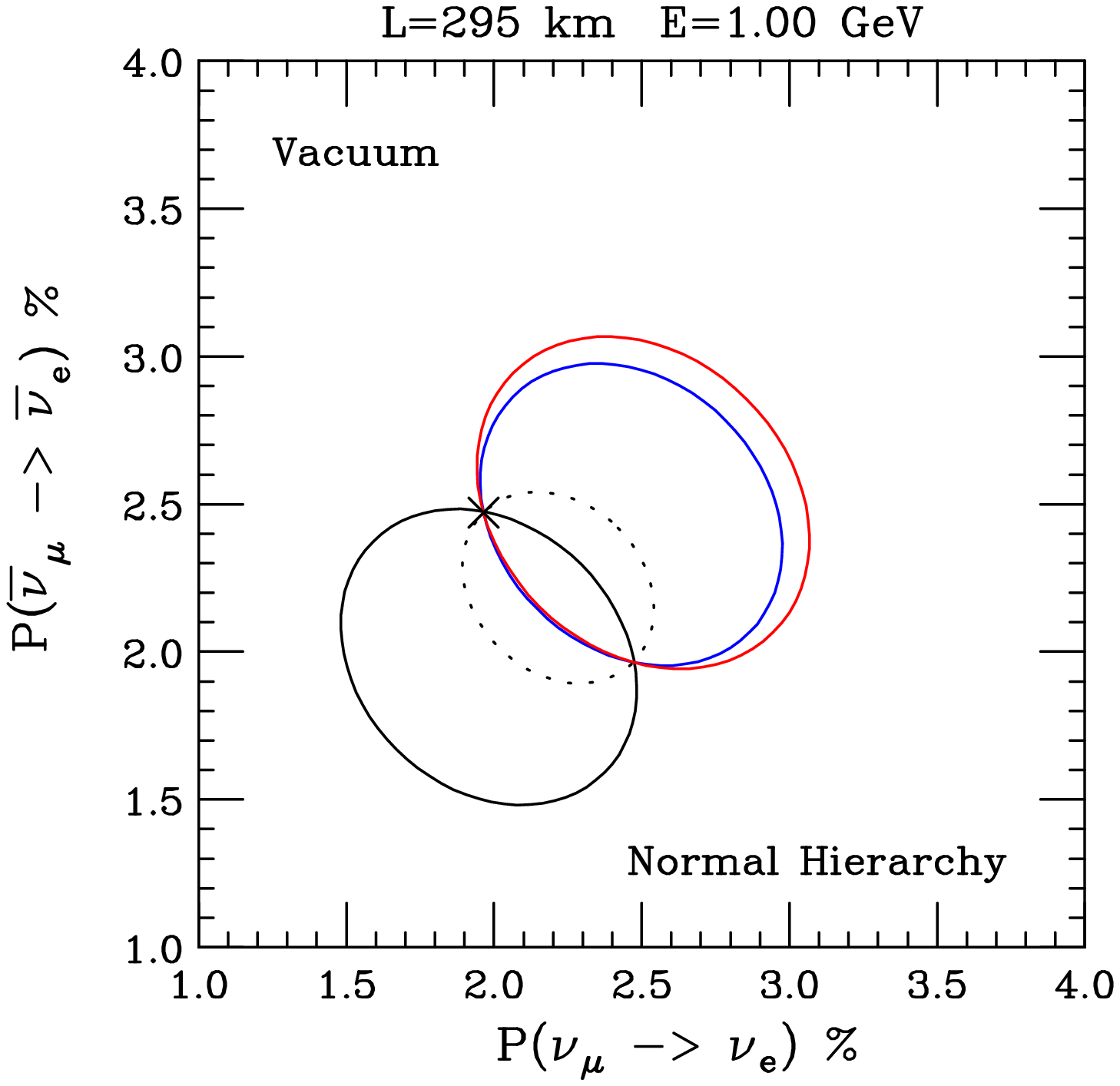}
\vspace{-8mm}
\end{center}
\caption{Left panel: 
Set of points in the $\sin^2 \theta_{13}$-$\sin^2 \theta_{23}$ plane which simultaneously give $P(\nu_{\mu} \to \nu_e) \approx 0.02$ and $P(\bar{\nu}_{\mu} \to \bar{\nu}_e) \approx 0.025$ 
 in vacuum, for $E_\nu=1$ GeV and $L=295$ km. Each point in the curve corresponds to a different value of $\delta$. 
Larger stars on the curve mark the points corresponding to the $\theta_{23}$ and $\theta_{13}$ intrinsic degeneracy solutions. Small stars indicate values of $\delta$ in steps of $10^\circ$, from $40^\circ$ to $140^\circ$. Notice the accumulation of points $\delta=80^\circ$, $90^\circ$, and $100^\circ$ near the tip of the ``tusk''.  
Right panel: 
The bi-probability plot in $P(\nu_{\mu} \to \nu_e)$ vs. $P(\bar{\nu}_{\mu} \to \bar{\nu}_e)$ space. The points with large stars on the curve in the left panel correspond to the ellipses with the same color in the right panel.  The dotted curve is the smallest ellipse that can be drawn through these points. 
}
\label{fig:bi-P-plot}
\end{figure}

\subsubsection{The Appearance Degeneracy and the relationship to the Intrinsic Degeneracies}

If we solve Eq.~(\ref{Pemu-matter}) for $\theta_{23}$ and $\theta_{13}$ by eliminating $\delta$ at a given neutrino energy $E$ and a baseline $L$, a curve on $\sin^2 \theta_{13}$ vs. $\sin^2 \theta_{23}$ plane results. An example of such a curve is drawn in vacuum in the left panel of Fig.~\ref{fig:bi-P-plot} by varying $\delta$ for a setup with $L=295$ km and a neutrino energy of 1 GeV. That is, if we set up the problem so that we obtain solutions for $\theta_{23}$ and $\theta_{13}$ by measurement of $P \equiv P(\nu_{\mu} \to \nu_e)$ and $\bar{P} \equiv P(\bar{\nu}_{\mu} \to \bar{\nu}_e)$ at a certain value of energy, we have solutions on any points on the curve; the degeneracy is continuous. In other words, because of the freedom of adjusting $\theta_{13}$ and $\theta_{23}$ to reproduce the measurement points ($P$, $\bar{P}$), the solutions are in fact not only at the discrete points but on a continuous line parameterized by $\delta$ e.g., $\theta_{23}$ expressed as a function of $\theta_{13}$ as in the left panel of Fig.~\ref{fig:bi-P-plot}.

To reveal the features of the appearance degeneracy and to understand its relationship to the $\theta_{23}$ and $\theta_{13}$ intrinsic degeneracies, let us do the following exercise.  Suppose that the true values of the parameters are at $\sin^2 \theta_{23} = 0.4$, $\sin^2 \theta_{13}=0.02$, and $\delta=40^\circ$ as indicated by the black star in the left panel of Fig.~\ref{fig:bi-P-plot}. If we know $\theta_{13}$ exactly we have a clone solution at $\sin^2 \theta_{23}^{(2)} = 0.5$, $\sin^2 \theta_{13}^{(2)}=0.02$, and $\delta^{(2)}=140^\circ$, as indicated by the blue star. This is nothing but an example of the $\theta_{23}$ intrinsic degeneracy. 
Notice that $\delta^{(2)}=\pi - \delta$ as it should be in vacuum. Whereas if we know $\theta_{23}$ exactly we have the third solution shown by the red star in Fig.~\ref{fig:bi-P-plot} at $\sin^2 \theta_{23}^{(3)} = 0.4$, $\sin^2 \theta_{13}^{(3)}=0.0255$, and $\delta^{(3)}=146^\circ$, an example of the $\theta_{13}$ intrinsic degeneracy.\footnote{
A similar description with figures like Figs.~\ref{fig:bi-P-plot} and \ref{fig:int-edep} of how appearance and disappearance measurements can solve the $\theta_{23}$ disappearance ``octant'' degeneracy appeared in~\cite{Hiraide:2006vh}. The correlation between $\theta_{13}$ and $\theta_{23}$, which has been noticed since early times, \eg\ in \cite{Huber:2002mx}, seems to reflect at least partly the effect of the ``tusk'' shaped correlation displayed in the left panel of Fig.~\ref{fig:bi-P-plot}. 
}

The fact that each degenerate solution is able to reproduce the measured quantities ($P\approx 0.02$, $\bar{P}\approx0.025$ in this particular case) can be easily seen if we use the bi-probability plot in $P - \bar{P}$ space~\cite{Minakata:2001qm}. In the right panel of Fig.~\ref{fig:bi-P-plot} the three  bi-probability ellipses corresponding to the three degenerate solutions (the true point and two fake clones) are drawn by using the same color as used in the left panel.\footnote{
The reader may wonder about the meaning of dotted ellipse in the right panel of Fig.~\ref{fig:bi-P-plot}. It is the special case with the smallest size of the ellipse. There is a unique way to draw the minimum size ellipse passing through the measurement point ($P$, $\bar{P}$). Namely, it is to place the ellipse so that its edge just touches to the point ($P$, $\bar{P}$) as marked by the cross in the right panel of Fig.~\ref{fig:bi-P-plot}. Since the upper-left edge of the ellipse always correspond to $\delta = 90^\circ$ the solution must correspond to this value of $\delta$. Since this ellipse is unique by definition, there is no degeneracy in this case. Therefore, the point $\delta = 90^\circ$ must be at the tip of the ``tusk'', as shown by the red star in the left panel of Fig.~\ref{fig:bi-P-plot}. 
}

\subsubsection{$\theta_{23} - \theta_{13} - \delta$ Appearance Degeneracy is fragile}
\label{sec:fragility}

\begin{figure}[b]
\vspace{-2mm}
\begin{center}
\includegraphics[width=0.52\textwidth]{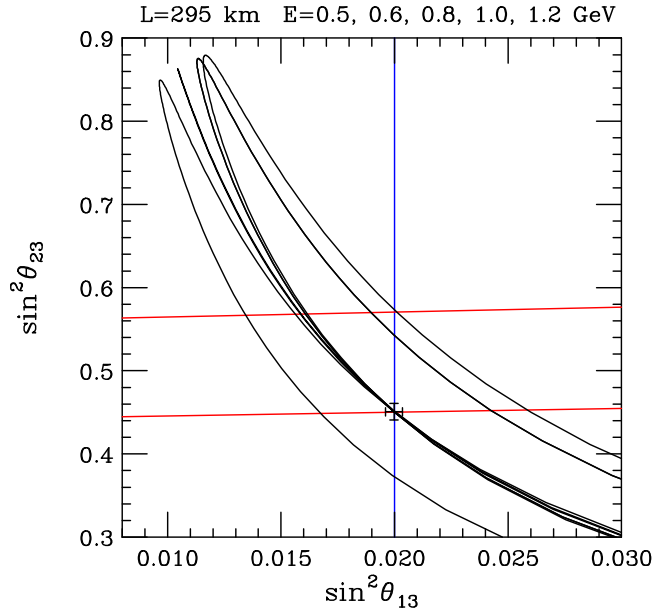}
\vspace{-6mm}
\end{center}
\caption{The appearance measurement degeneracy for four different neutrino energies, from left to right, 0.50, $ \sim 0.6$, 0.80 and 1.00 GeV using a baseline of 295km. The true input value (black cross) corresponds to $\sin^2 \theta_{23}=0.45$, $\sin^2 \theta_{13}=0.020$ and $\delta=30^\circ$. The black curve for $E\sim 0.6 (0.58)$~GeV is special because it corresponds to the VOM for this baseline, $\Delta_{31}=\pi/2$. Therefore, the bi-probability ellipses are squashed to a line, and the degeneracy folds over upon itself. This figure clearly shows that this degeneracy is ``fragile'' in the sense of being dependent on $L/E$. The vertical blue line is given by $\sin^2 \theta_{13}=0.020$ and the, almost horizontal, red lines show the solutions corresponding to $\sin^2 2 \theta_{\mu \mu} = 0.986$, see next subsection. Note, that the appearance degeneracy line for an energy of 1.0 GeV passes through both intersection points of the fixed $\sin^2 \theta_{13}$ and  fixed $\sin^2 2 \theta_{\mu \mu} $ constraints.
}
\label{fig:int-edep}
\end{figure}

It has been recognized that the $\theta_{13}$- and the $\theta_{23}$-intrinsic degeneracies are ``fragile'' in the sense that the position of the fake solutions is energy dependent so that spectrum measurements can be used to rule them out. It is worth noting that this fragility continues to be true for the $\theta_{23} - \theta_{13} - \delta$ appearance degeneracy. In Fig. \ref{fig:int-edep}, the position of the appearance degeneracy is shown for different values of $L/E$. As can be seen, the position of the appearance degeneracies changes as the value of $L/E$ is varied. However, there is one common solution for all four values $L/E$ of the experiment, which is of course the unique true solution. Thus, spectral information will be particularly valuable for eliminating the fake solutions provided there is ample statistics in each of the energy bins. This may be contrasted to the feature of disappearance ``octant'' degeneracy (see Sec.~\ref{sec:disappearance}) for which the clone solution is $L/E$ independent.
%

To close this subsection we would like to emphasize that the $\theta_{23} - \theta_{13} - \delta$ appearance degeneracy, for the three parameters $\theta_{23}$, $\theta_{13}$ and $\delta$,
is a continuous degeneracy  of the combined $\nu_e$ and $\bar{\nu}_e$ appearance probabilities only.

\subsection{Disappearance Channels and the $\theta_{23}-\theta_{13}$ Disappearance Degeneracy}
\label{sec:disappearance}

Reactor electron antineutrino disappearance experiments with baselines appropriate to observer atmospheric oscillations, such as Daya Bay~\cite{An:2013zwz}, RENO~\cite{fortheRENO:2013lwa} and Double Chooz~\cite{Abe:2012tg} experiments, have values of $L/E$ $\sim$ 0.5~km/MeV. They measure the oscillation probability $P(\bar{\nu}_e \rightarrow \bar{\nu}_e)$
\begin{eqnarray}
P(\bar{\nu}_e \rightarrow \bar{\nu}_e) = 1 - \sin^2 2\theta_{13} \sin^2 \left( \frac{\Delta m^2_{ee} L}{ 4 E } \right) + {O}(\Delta_{21}^2) \, ,
\label{Pee}
\end{eqnarray}
where $\Delta m^2_{ee}$ is the electron neutrino weighted average of $\Delta m^2_{31}$ and $\Delta m^2_{32}$~\cite{Nunokawa:2005nx}. In principle there is an octant degeneracy here for $\theta_{13}$ since the measurement of $\sin^22\theta_{13}$ does not allow to distinguish $\theta_{13}$ from $\pi/2 - \theta_{13}$. However, the Super-Kamiokande (Super-K) atmospheric neutrino results \cite{Wendell:2010md} ($|U_{\mu3}|^2 = \cos^2\theta_{13}\sin^2\theta_{23} \approx 1/2$) imply that $\theta_{13}$ is relatively small (and therefore in the first octant). This results in an unambiguous,  precise measurement of $\theta_{13}$
\begin{eqnarray}
\sin^2 \theta_{13} \approx 0.023.
\end{eqnarray}

For the muon neutrino disappearance experiments at the atmospheric baseline divided by neutrino energy, $L/E\sim 500$~km/GeV, such as K2K~\cite{Ahn:2006zza}, MINOS~\cite{Adamson:2011ig}, T2K~\cite{Abe:2014ugx} and NO$\nu$A~\cite{Ayres:2004js}, the muon neutrino survival probability is given by
\begin{eqnarray}
P(\nu_\mu \rightarrow \nu_\mu) = 1- \sin^2 2 \theta_{\mu \mu} \sin^2 \left( \frac{\Delta m^2_{\mu \mu} L}{ 4 E } \right) 
+ {O}(\Delta_{21}^2) \, ,
\end{eqnarray}
where $\Delta m^2_{\mu\mu}$ is the muon neutrino weighted average of $\Delta m^2_{31}$ and $\Delta m^2_{32}$~\cite{Nunokawa:2005nx}, and 
\begin{eqnarray}
\sin^2 2 \theta_{\mu \mu} \equiv 4 |U_{\mu3}|^2(1- |U_{\mu3}|^2) = 4\cos^2 \theta_{13} \sin^2 \theta_{23} (1- \cos^2 \theta_{13} \sin^2 \theta_{23}) \, .
\end{eqnarray}
Matter effects are very small in this channel (except maybe for some neutrino factory setups), and are ignored here.

\begin{figure}[htbp]
\begin{center}
\vspace{2mm}
\includegraphics[width=0.433\textwidth]{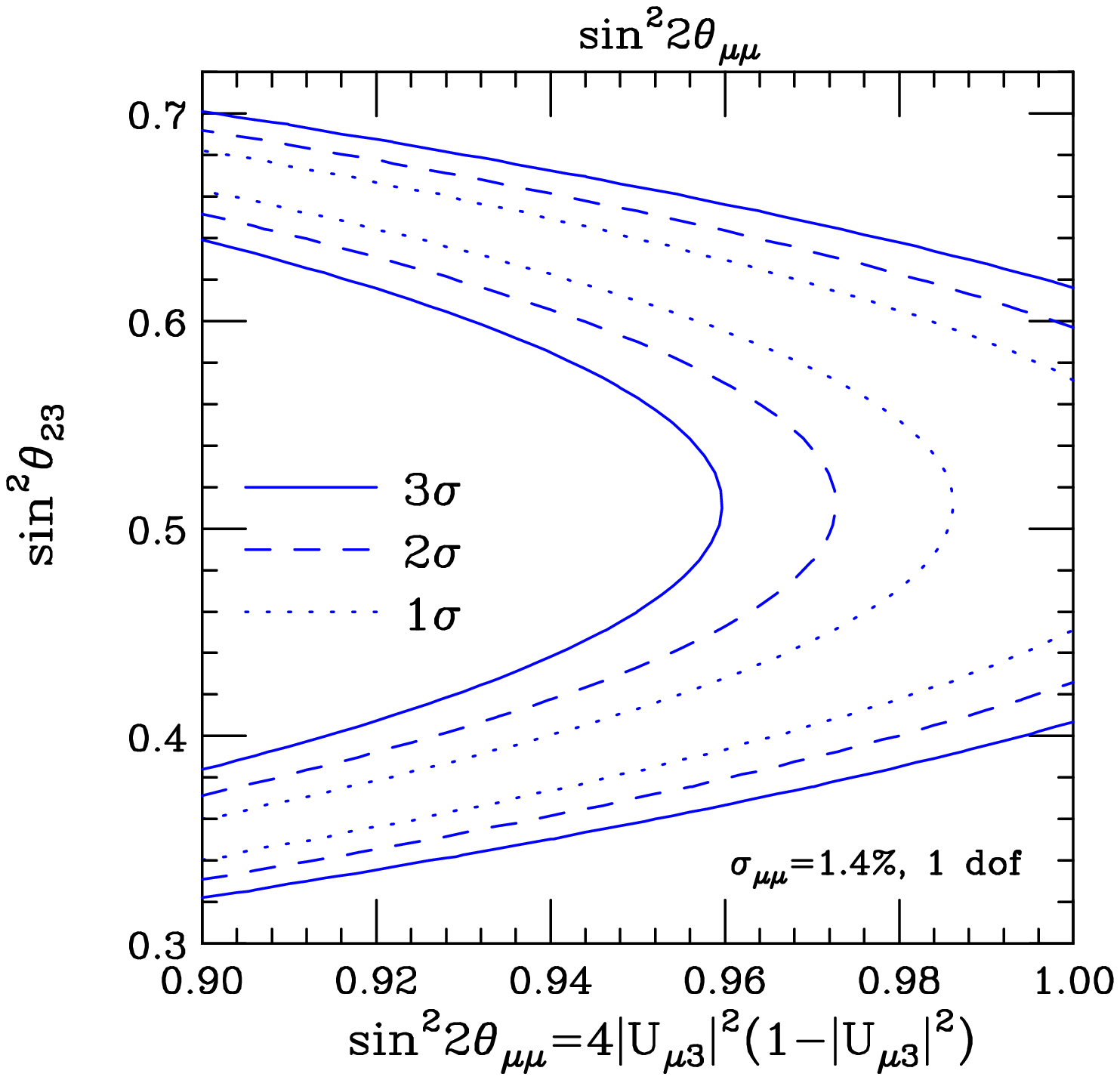}
\includegraphics[width=0.42\textwidth]{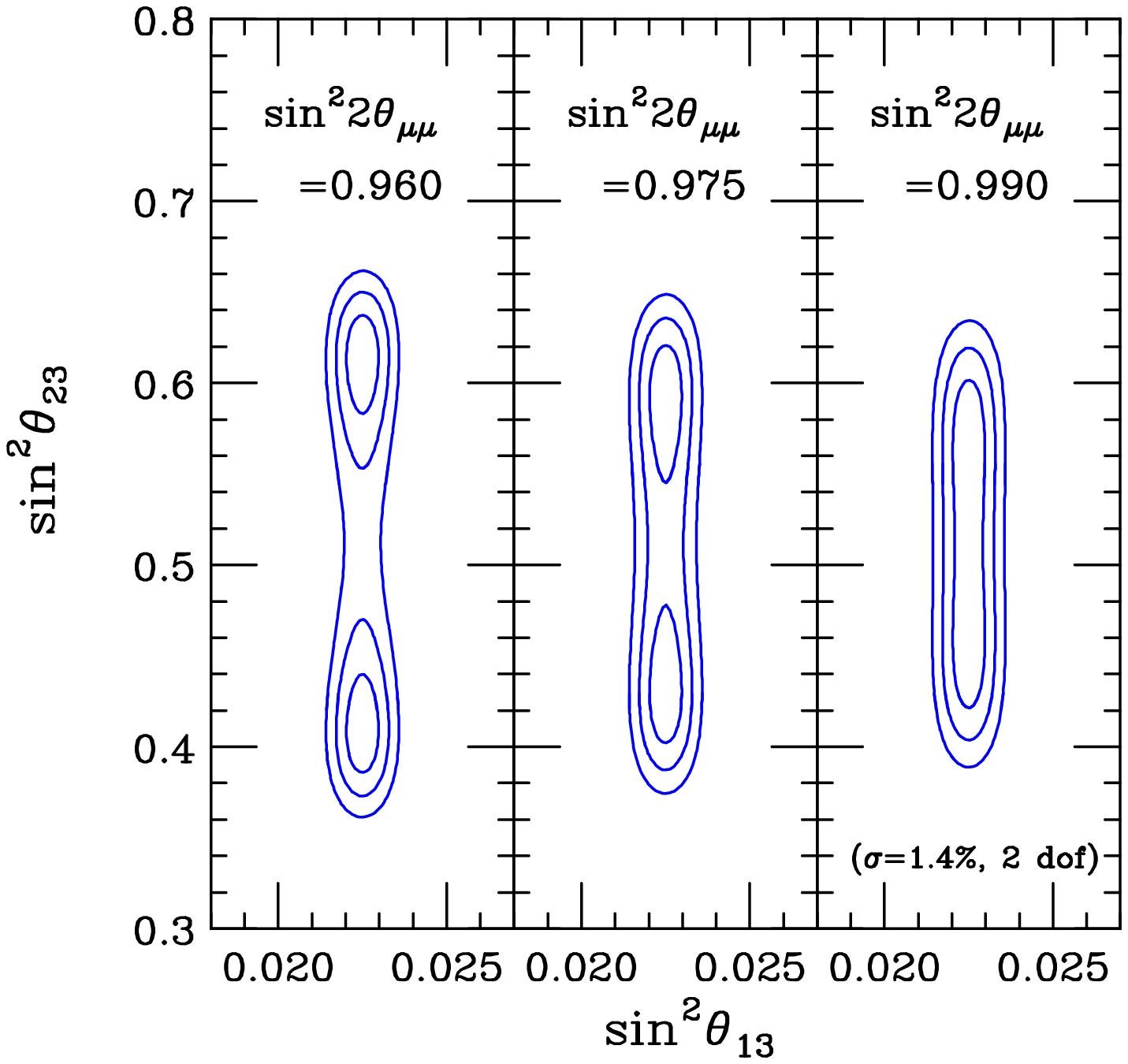}
\vspace{-2mm}
\end{center}
\caption{ Left panel: Contours for the $\chi^2$ distribution in the $\sin^2 2\theta_{\mu \mu}$-$\sin^2 \theta_{23}$ plane, where $\chi^2 \equiv (4\cos^2 \theta_{13}\sin^2 \theta_{23}(1-\cos^2 \theta_{13}\sin^2 \theta_{23})-\sin^2 2\theta_{\mu \mu})^2/(\sigma_{\mu\mu} \sin^2 2\theta_{\mu \mu})^2$, $\sin^2 \theta_{13}=0.023$ and $\sigma_{\mu\mu}=1.4\%$. Right panel: Contours for the same $\chi^2$ in the $\sin^2\theta_{13}$ and $\sin^2 \theta_{23}$ plane, for different values of $\sin^22\theta_{\mu\mu}$ as indicated in each sub-panel. With the assumed uncertainty, there are two distinct allowed bands for $\sin^2\theta_{23}$ for values of  $\sin^2 2\theta_{\mu \mu} < 0.96$, whereas the two bands start to merge for $\sin^2 2\theta_{\mu \mu} >0.96$. Note the small upward shift with respect to the line $\sin^2\theta_{23}=0.5$ caused by the non-zero value of $\sin^2 \theta_{13}$. In both panels, the different lines correspond to different confidence levels as indicated in the legend. Note that the left panel corresponds to 1 d.o.f. while the right panel is obtained for 2 d.o.f.
}
\label{fig:muchisq}
\end{figure}

For relatively small values of $\theta_{13}$, the fate of the determination of $\sin^2 \theta_{23}$ depends very much on how  close $\theta_{23}$ is to the maximal value. In Fig \ref{fig:muchisq}, we have plotted the $\chi^2$ of  $\sin^2 2\theta_{\mu \mu}$ as a function of $\sin^2 \theta_{23}$ assuming an uncertainty of 1.4\% for the labeled various central values for $\sin^2 2\theta_{\mu \mu}$.  Using this uncertainty the two regions start to merge when $\sin^2 2\theta_{\mu \mu} >0.96$ and the determination of $\sin^2 \theta_{23}$ from the $\nu_\mu$-disappearance measurements is significantly degraded.  (The critical value which separates the two regions, of course, will depend on the actual accuracy of the measurement.)

A measurement of $\sin^2 2\theta_{\mu\mu}$ gives two distinct values of $\sin^2 \theta_{23}$ given by
\begin{eqnarray}
\sin^2 \theta^{(1)}_{23} =  \sin^2 \theta_{\mu \mu}/\cos^2 \theta_{13} &\approx& \sin^2 \theta_{\mu \mu}(1 +  \sin^2\theta_{13}) \, ,
\nonumber \\
\sin^2 \theta^{(2)}_{23} = \cos^2 \theta_{\mu \mu}/\cos^2 \theta_{13} &\approx& \cos^2 \theta_{\mu \mu} (1+  \sin^2\theta_{13}) \,  ,
\label{octant-sol}
\end{eqnarray}
using the convention that $\theta_{\mu \mu} \leq \frac{\pi}{4}$, i.e. $\sin^2\theta_{\mu \mu} \leq \frac{1}{2}$. Note, that $\theta^{(2)}_{23}$ is {\it always} in the second octant and for \emph{nearly all} values of  $ \theta_{\mu \mu} $,  $\theta^{(1)}_{23}$ is  in the first octant. However, if 
\begin{eqnarray}
\sin^2 \theta_{\mu \mu} > \frac{1}{2} \cos ^2 \theta_{13} \nonumber
\end{eqnarray}
then $\theta^{(1)}_{23}$ is  {\it also} in the second octant. This new feature of the $\theta_{23}$ ``octant'' degeneracy only occurs if $\theta_{23}$ is very close to maximal and for the observed non-zero value of $\theta_{13}$.
 
 For $\sin^2 2\theta_{23} \simeq \sin^2 2\theta_{\mu\mu}  \gsim 0.96$, 
the two allowed regions of $\sin^2\theta_{23}$ merge to a unique one which is extended to both the first and the second octants of $\theta_{23}$. Exactly where this occurs depends on the systematic errors used in the disappearance measurement.
An example is shown in the right panel of Fig.~\ref{fig:muchisq}. In this merged region, information on the value of $\sin^2 \theta_{23}$ from the appearance channels will be particularly useful.

What is currently known about  $\sin^2 2 \theta_{\mu \mu}$? The recent $\nu_\mu$-disappearance measurement by T2K  reported $\sin^2 2 \theta_{\mu \mu} \gsim 0.97$ at 90\% CL (1 d.o.f.)~\cite{Abe:2014ugx,Minamino-KEK}. Thus, it appears that nature has chosen to live in this merged region, on which we focus in the following discussion. 

To close this subsection we would like to emphasize that this $\theta_{23} - \theta_{13}$ disappearance degeneracy (or ``octant'' degeneracy),   for the two parameters $\theta_{23}$ and $\theta_{13}$, is a continuous degeneracy of the $\nu_\mu$ disappearance probability only.

\subsection{Features of the general $\theta_{23} - \theta_{13} - \delta$ continuous degeneracy}
\label{sec:continuous-P}

In the merged region, $\sin^2 2\theta_{\mu \mu} \gsim 0.96$, we face with two kinds of continuous degeneracies: the $\theta_{23} - \theta_{13}-\delta$ appearance degeneracy and the $\theta_{23}- \theta_{13}$ disappearance degeneracy. In this subsection we discuss some of the interesting features of these degeneracies and their intersection.

\begin{figure}[b]
\vspace{2mm}
\begin{center}
\includegraphics[width=0.46\textwidth]{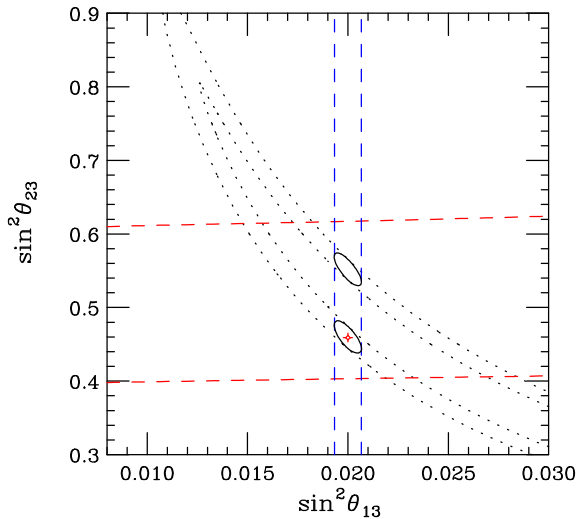}
\end{center}
\caption{Allowed confidence regions in the $\sin^2\theta_{13}$-$\sin^2\theta_{23}$ plane at the 2 $\sigma$ CL (2 d.o.f.), using different oscillation channels,  for an experiment with a baseline of 295~km and a (monochromatic) neutrino energy of 0.8~GeV (\ie, $ \Delta_{31} \sim 3 \pi/8$). The allowed region for the appearance $\nu_\mu \rightarrow \nu_e$ and $\bar{\nu}_\mu \rightarrow \bar{\nu}_e$ measurements is given by the dotted black bands. The vertical blue and (almost) horizontal red bands indicate the regions allowed by the $\nu_e$ and $\nu_\mu$ disappearance measurements constraints on $\sin^2 2\theta_{13}$ and $\sin^2 2\theta_{\mu \mu}$. Finally, the solid black ellipses are the overlap regions for these three types of measurements.   }
\label{fig:23-region}
\end{figure}

In Fig.~\ref{fig:23-region}, the allowed regions in the $\sin^2\theta_{23}-\sin^2 \theta_{13}$ plane are shown for 2~$\sigma$ CL (2 d.o.f). The vertical blue band comes from the $\bar{\nu}_e$ disappearance measurement, the almost horizontal red band corresponds to the merged first and second octant solutions for the $\nu_\mu$ disappearance measurements and the dotted black band is the $\nu_\mu \rightarrow \nu_e$ and $\bar{\nu}_\mu \rightarrow \bar{\nu}_e$ appearance measurements.  In all cases, the uncertainties on the measurement are assumed to be of 1.4\%, and are implemented as in Ref.~\cite{Minakata:2013eoa}.  
This figure clearly shows the continuous degeneracy in the $\sin^2\theta_{23}$ vs $\sin^2 \theta_{13}$ plane associated with the appearance and disappearance probabilities, as well as the overlapping regions between them. One of these overlapping regions is the true solution, while the other region is fake and will move as we vary the neutrino energy.\footnote{
These two solutions, which stem from the appearance degeneracy, could have been misunderstood as a consequence of the disappearance ``octant'' degeneracy, if the appearance and disappearance channels are analyzed simultaneously to obtain the allowed regions. This can be understood from Fig.~\ref{fig:int-edep} as well. It can be seen in Fig.~\ref{fig:int-edep} that, at $\sin^2 \theta_{13}=0.02$, the appearance degeneracy curve with E=1.0 GeV has two allowed solutions for $\sin^2\theta_{23}$: the true one (indicated by a black cross), plus an appearance clone solution which overlaps with the disappearance (octant) clone. However, such overlap occurs only for isolated values of the neutrino energy. 
}

The assumed true input values for the oscillation parameters in this case are
\begin{eqnarray}
\sin^2 \theta_{23}=0.45, \quad \sin^2 \theta_{13}=0.020 \quad {\rm and} \quad \delta=30^\circ \, . 
\end{eqnarray}
This is represented in this figure by the black solid lines with the red cross in the center.  The other region which also satisfies all the measurements is located at a larger 
value of $\sin^2 \theta_{23}$. Its exact position will depend on the value of $L/E$ for the experiment, and it will be located at a value of $\delta^{(2)} = \pi - \delta$. This second solution will move up and down within the vertical blue band between the two sets of horizontal red lines depending on the neutrino energy, as it was shown in Fig. \ref{fig:int-edep}.  Thus, spectral information would be very powerful
in removing this degeneracy, provided the statistics is sufficient in several well-defined energy bins.

\section{Experimental setups}
\label{sec:exp}

Four experimental setups are considered in this work. We believe these are representative of four different types of neutrino oscillation experiments, according to their values of $L/E$. As it stands, the nature of the four settings to be examined is not intended for a performance comparison between the different setups but to illuminate their characteristic features based on different physical principles: 

\begin{enumerate}

\item \textit{Narrow band beams operating at the first VOM}, 
$\Delta_{31} \sim \pi/2$. The beam is aimed to the detector at an off-axis location, so that the flux is very narrow in energy, centered around the first VOM. This is the case of T2K~\cite{Abe:2011ks} and NO$\nu$A~\cite{Ayres:2004js}, for instance. In this work we will consider 
an upgrade of T2K, which is usually referred to T2HK~\cite{Abe:2011ts}.
It will use the same beamline as T2K uses, aiming instead at a 560~kt fiducial volume water \v{C}erenkov detector Hyper-Kamiokande (Hyper-K) to be placed at the same distance (295~km) and off-axis angle ($2.5^\circ$) as Super-K. 

\item \textit{Wide band beams (WBB) operating around the first VOM}, \textit{i.e.}, $\Delta_{31} \sim (2\pm1)\pi/4$. These experiments are performed on-axis. Therefore, the beam flux is much wider as in the previous case and as a consequence they observe not only the first VOM but also some regions above and below it. The main advantage of this type of experiments is that the oscillation pattern is much better reconstructed, and the statistics is much larger since the detector is placed on-axis. Examples of these type of experiments are LBNE~\cite{Akiri:2011dv,CDR,Adams:2013qkq} and LBNO~\cite{Stahl:2012exa}. In this work, we will consider the LBNE experiment, which consists of a 1.2~MW beam and a 34~kt Liquid Argon (LAr) detector placed underground, at a baseline of 1300~km from the source.

\item \textit{Neutrino beams operating below VOM}, \textit{i.e.}, $\Delta_{31} < \pi/2$. The Neutrino Factory (NF) setups traditionally considered in the literature would operate in this regime, see for instance Ref.~\cite{Choubey:2011zzq}. More recently, lower energy versions have also been proposed, see for instance~\cite{Agarwalla:2010hk,FernandezMartinez:2010zza,Christensen:2013va}, which operate in a regime much closer to the first VOM. In this work we will consider a NF setup operating below VOM, such as the IDS-NF setup~\cite{idsnfweb}, with a baseline of 2000~km and a parent muon neutrino energy of $E_\mu=10$~GeV~\cite{Agarwalla:2010hk}. For this setup, we consider a 100~kt Magnetized Iron Neutrino Detector (MIND).

\item \textit{Neutrino beams operating at the second VOM,} \textit{i.e.}, $\Delta_{31} = 3\pi/2$. This is the case of the recently proposed ESS$\nu$SB facility in Europe~\cite{Baussan:2012cw,Baussan:2013zcy}. At the second VOM, the size of the $\delta$-dependent interference term between the atmospheric and the solar terms is a factor of $\sim 3$ larger than that at the first VOM, which would lead to higher sensitivity to $\delta$. The favorable feature is utilized in this and in the earlier proposals, \eg , in~\cite{Marciano:2001tz,Diwan:2002xc,Ishitsuka:2005qi,Coloma:2011pg}. 
Here, we consider one of the setups within the ESS$\nu$SB proposal, which consists of a 500~kt fiducial mass water \v{C}erenkov detector placed at 540~km from the source. As for the beam, we will consider a 5~MW beam produced using 2.5~GeV protons.
\end{enumerate}

Table~\ref{tab:setups} summarizes the main features of the setups considered in this work. The different columns indicate the baseline, neutrino flux peak, beam power per year (or number of useful muon decays, in the case of the IDS-NF), detector size and data taking period for neutrinos and antineutrinos. 
\begin{table}
\begin{center}
\renewcommand\tabcolsep{6pt}
\renewcommand\arraystretch{1.5}
\begin{tabular}{l|cccccc}
  & $L$ (km) & Detector (kt) & Beam Power &  $E_p$ (GeV) &  Flux peak & $(t_\nu, t_{\bar\nu})^\dagger\times 10^7 s$ \\ \hline
LBNE  & 1300 &  LAr - 34 & 1.2~MW & 120  & 3 GeV & (8.25, 8.25) \\  
T2HK  & 295 &  WC - 560 & 0.75~MW & 30 & 0.6 GeV & (3, 7) \\  
ESS$\nu$SB  & 540 & WC - 500 & 5~MW & 2.5 & 0.3 GeV & (3.4, 13.6) \\  
IDS-NF  & 2000 &  MIND - 100 & $10^{21}$ $\mu^{\pm}$/$10^7$ sec & NA & 6 (9) GeV & (10, 10)$^\ddagger$ \\   \hline   
\end{tabular}
\caption{Main features of the experimental setups simulated in this work. The different columns indicate the distance to the detector, the detector technology, its mass, the beam power (or the number of useful muon decays per year in the case of the IDS-NF), the energy at which the neutrino flux peaks and the running time (in units of $10^7$ seconds) for $\nu$ and $\bar\nu$ modes. Here, LAr stands for Liquid Argon, WC for Water \v{C}erenkov, and MIND for Magnetized Iron Neutrino Detector. Note also that in the case of the IDS-NF the energy for the flux peak for both $\nu_e$ and $\bar\nu_\mu$ (in parenthesis) are separately indicated, since they are different. For the number of events in each oscillation channel see Table~\ref{tab:events}. 
\newline$^\dagger$ 
Note that each experiment assumes a different number of operating seconds per calendar year. LBNE and ESS$\nu$SB assume $\sim 1.7\times 10^7$ operating sec/year, while T2HK and IDS-NF assume $1.0 \times10^7$ operating sec/year. This implies that the running time for all the experiments considered in this work is expected to be 10 calendar years. 
\newline$^\ddagger$While for conventional beams the running time is split between neutrino ($\pi^+$-focusing) and antineutrino ($\pi^-$-focusing) modes, the IDS-NF setup assumes that both $\mu^+$ and $\mu^-$ would run at the same time in the decay ring. The total number of muon decays per year would be equally split between the two polarities in this case.
\label{tab:setups} }
\end{center}
\end{table}
Technical details used to simulate each setup, as well as the number of events for each oscillation channel, can be found in Appendix~\ref{app:sim}, together with a brief explanation of the $\chi^2$ implementation, the inclusion of systematic errors in our analysis, the values of the oscillation parameters and the marginalization procedure.

\section{Appearance and disappearance measurements of $\mathbf{\theta_{23}}$}
\label{sec:relative}

Given the understanding of mutual roles played by the $\nu_{e}$ appearance and the $\nu_{\mu}$ disappearance channels in resolving the general $\theta_{23} - \theta_{13} - \delta$ degeneracy, we now discuss in more closely their relative importance in accurate measurement of $\theta_{23}$, and how it depends on the systematic errors. The answer to this question depends on the particular setup under consideration. Therefore, we focus on the two typical cases, T2HK/LBNE and IDS-NF in Secs.~\ref{sec:s23-T2HK} and \ref{sec:s23-NF}, respectively. We will also comment briefly on the results for the ESS$\nu$SB experiment. In Sec.~\ref{sec:23-delta} we discuss the relationship between errors of $\sin^2\theta_{23}$ and $\sin \delta$ using the appearance measurement only.

We note that the issue of mutual role by the appearance and disappearance channels for $\theta_{23}$ determination has been addressed, \eg , in~\cite{Donini:2004iv,Donini:2005rn,Donini:2005db}. More generally, the sensitivity to $\theta_{23}$ has been discussed, though from somewhat different point of view, by many authors. The analyses from early to recent times include, for example, Refs.~\cite{Minakata:2004pg,GonzalezGarcia:2004cu,McConnel:2004bd,Huber:2004ug,Choubey:2005zy,Hiraide:2006vh, Agarwalla:2013ju,Chatterjee:2013qus,Machado:2013kya,Choubey:2013xqa} (see also Refs.~\cite{GonzalezGarcia:2012sz,Capozzi:2013csa,Tortola:2012te} for some results of global fits on the determination of $\theta_{23}$).

\subsection{Relative importance of appearance and disappearance channels for facilities sitting at the first VOM: T2HK and LBNE}
\label{sec:s23-T2HK}

In this subsection we discuss the precision that can be achieved for a measurement of $\sin^2\theta_{23}$ with the disappearance and/or the appearance channels for facilities sitting at the first VOM, and discuss their dependence with the systematic errors. 
The results of the analyses are presented in Fig.~\ref{fig:syst-dep-t2hk} and Fig.~\ref{fig:syst-dep-lbne} for the T2HK and LBNE setups, respectively. In both figures, we show the precision attainable for a measurement of $\sth$ as a function of the true value of $\sth$ itself, where left and right panels show the results under different assumptions for the systematic errors. In the left panel we use our default values ($\sim 5-10\%$), while more conservative values are assumed for the right panel ($\sim 10-15\%$). See Appendix~\ref{app:sim} for a more precise specification of the systematic errors and the way these are implemented in our analysis.\footnote{
The size of the expected systematic errors for future neutrino oscillation experiments can be a controversial subject, and is currently under study. For T2HK we have used the systematic uncertainties based on the HK LoI~\cite{Abe:2011ts}, and for LBNE they fall approximately within the same ballpark as those considered by the collaboration (see \eg, Ref.~\cite{Adams:2013qkq}). We note that the systematic errors for $\nu_e$ appearance measurement currently examined by the Hyper-K working group \cite{HK-sensitivity} are more optimistic than our default values. In this case the appearance sensitivity (solid line) in the left panel in Fig.~\ref{fig:syst-dep-t2hk} would supersede the disappearance one around $\sth \sim 0.49 $ and $\sth \sim 0.54$, recovering Region II.}

As it can be seen from the figures, not only the absolute sensitivity to $\sin^2\theta_{23}$ but also relative importance of $\nu_e$ appearance and $\nu_\mu$ disappearance measurement on the determination of $\theta_{23}$ depend very much on the size of the systematic errors. It is notable that not only this feature but also the absolute sensitivities to $\sin^2\theta_{23}$ (for our default systematic errors) are very similar between the T2HK and LBNE setups, despite their very different beam profiles and detector technologies. Yet, one observes that the LBNE setup would be more robust against the increase of systematic errors, probably because of its wide band beam.~\footnote{ Similar conclusions were obtained in Ref.~\cite{Coloma:2012ji} for different observables. } 

It is also quite noticeable the very different dependence of the results for the different oscillation channels with the true value of $\theta_{23}$. The results obtained through the appearance channels present very little dependence with this parameter for both LBNE and T2HK. On the other hand, the disappearance results present a very particular shape, with two very sharp peaks. This is due to the disappearance degeneracy (see Sec.~\ref{sec:G-degeneracy}): for values of $\theta_{23}$ close to maximal mixing the two solutions merge and the size of the confidence region is consequently worsened; the ``valley'' in the middle of the two peaks corresponds to the point where the two solutions lie exactly one on top of the other and therefore the precision is slightly improved. We can identify, generically, the following three regions with different characteristics. We note, however, that the exact locations of the boundaries between regions depend on the systematic and the statistical errors:

\begin{figure}
\begin{center}
\includegraphics[scale=0.45]{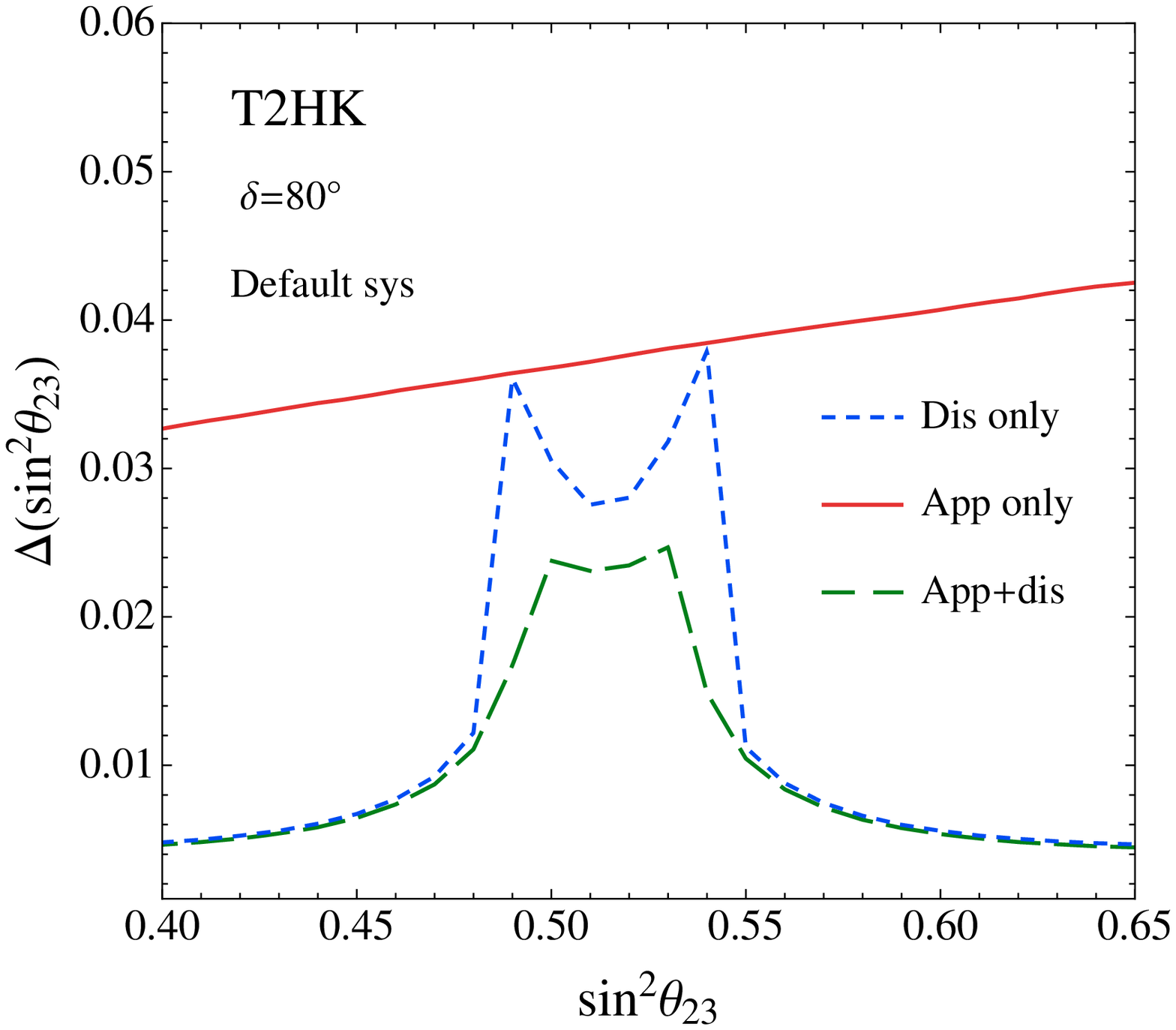} 
\includegraphics[scale=0.45]{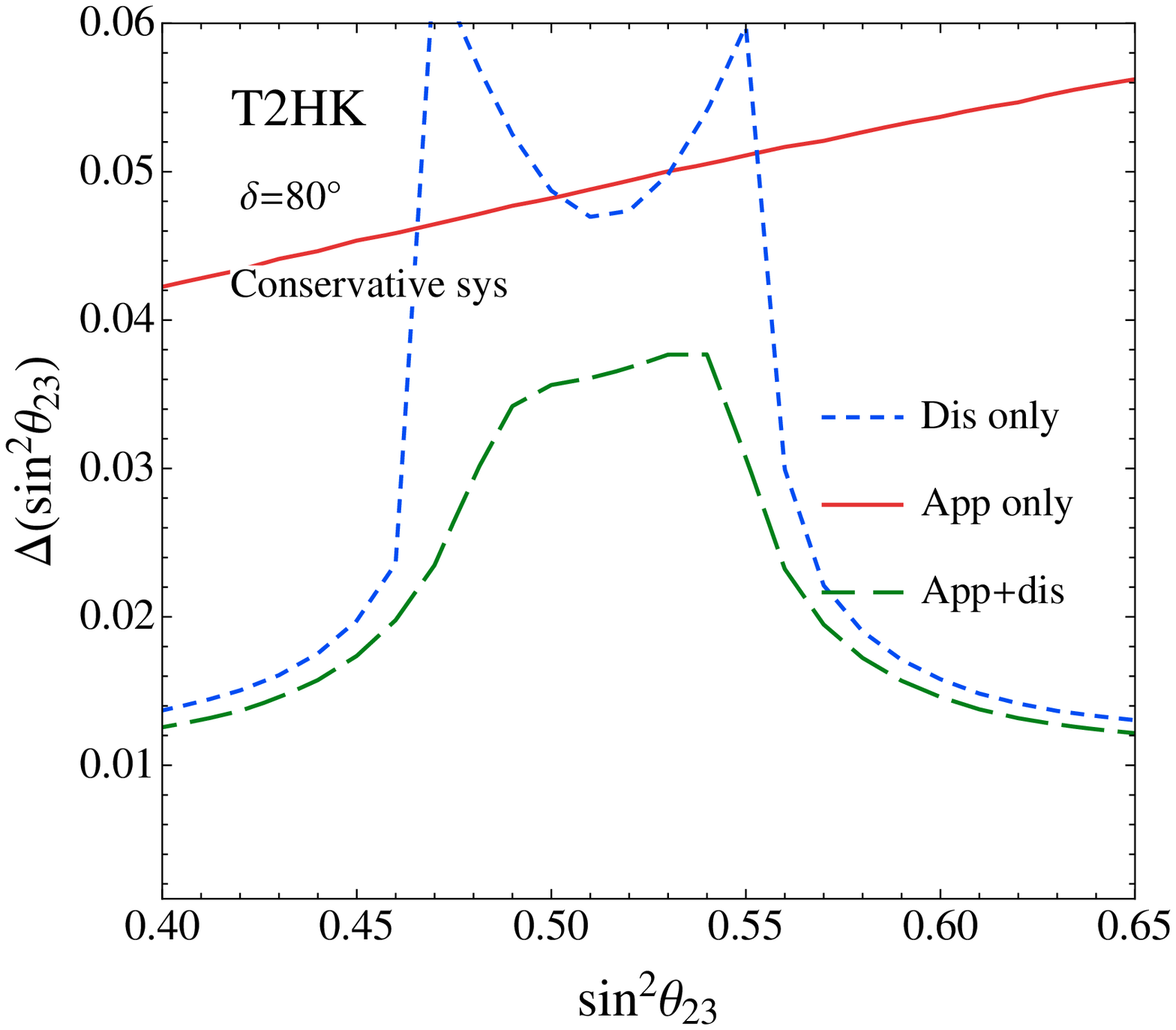} 
\caption{Expected precision for $\sin^2\theta_{23}$ at $1\sigma$ (1 d.o.f.) as a function of the true value of $\sin^2\theta_{23}$ for the T2HK setup. The left and the right panels correspond to our reference ($\sim 5-10\%$) and conservative ($\sim 10-15\%$) sets of systematic errors, respectively. See App.~\ref{app:sim} for more precise specification of the reference errors. The true value of $\delta$ is taken as 80$^\circ$.}
\label{fig:syst-dep-t2hk}
\end{center}
\end{figure}
%
\begin{figure}
\begin{center}
\begin{tabular}{cc}
\includegraphics[scale=0.45]{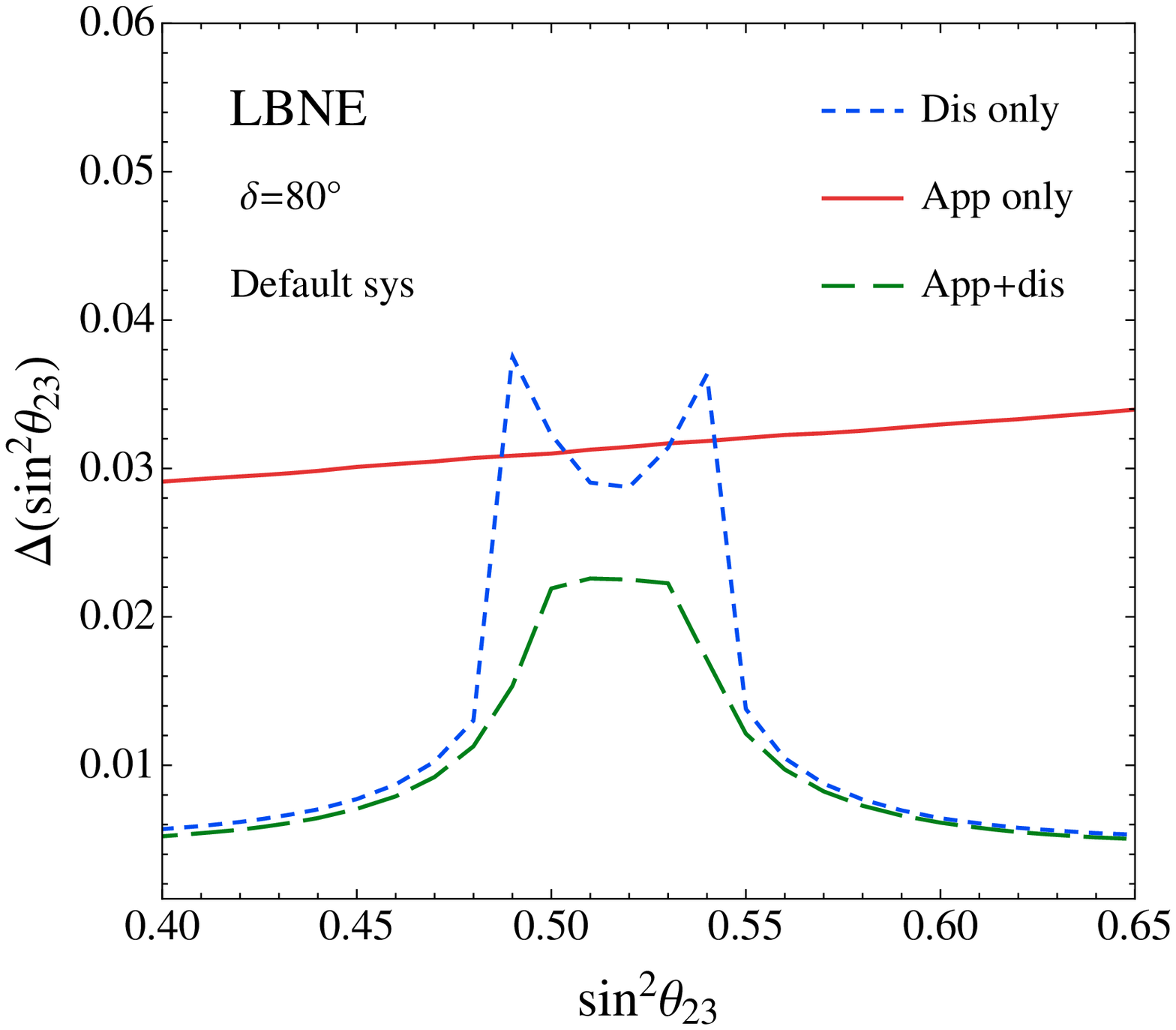} &
\includegraphics[scale=0.45]{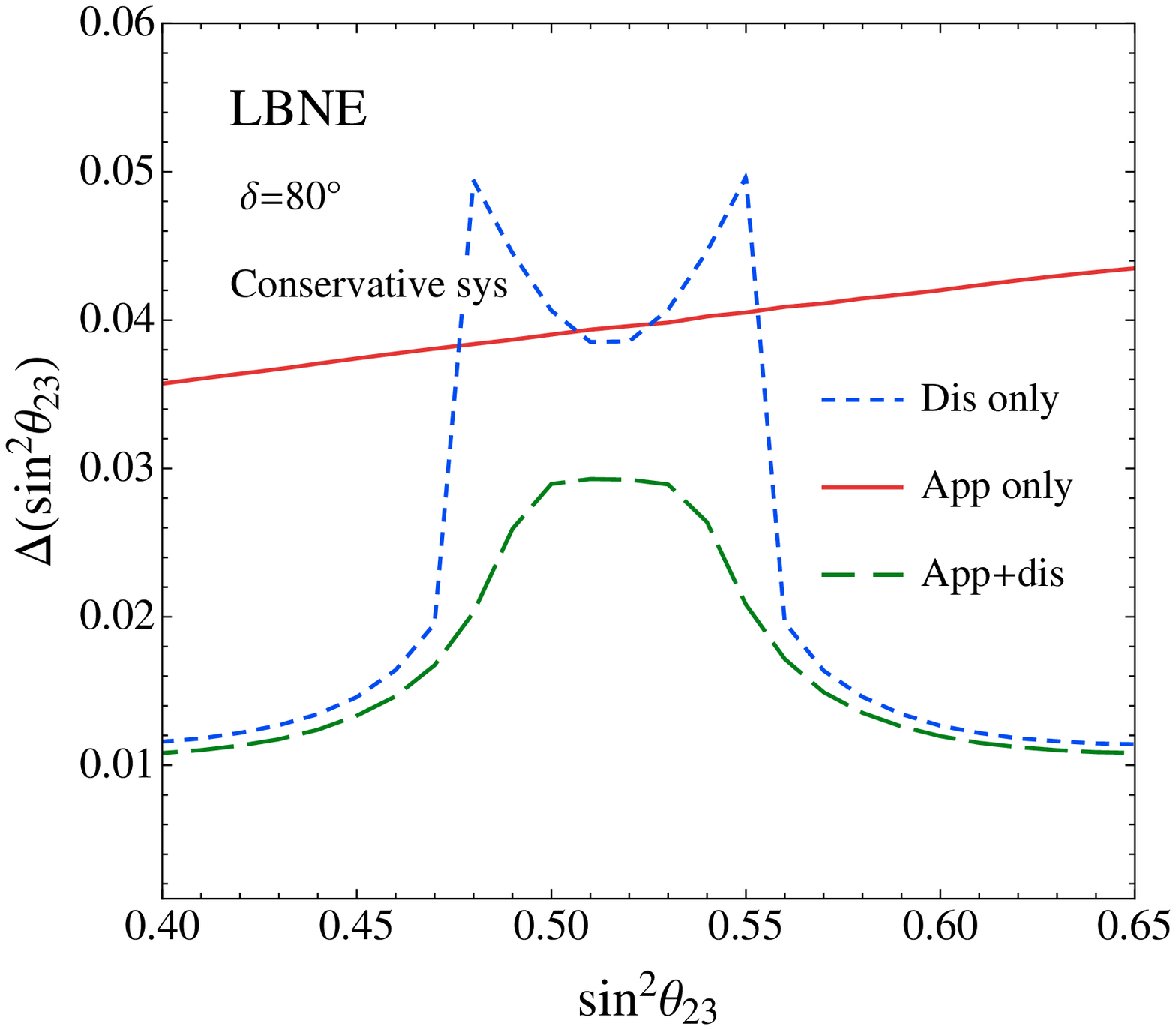}
\end{tabular}
\caption{Same as in Fig.~\ref{fig:syst-dep-t2hk}, but for the LBNE setup.}
\label{fig:syst-dep-lbne}
\end{center}
\end{figure}

\begin{description}
\item[Region I:] 
In the two regions where $\sin^2\theta_{23} \lsim 0.46$ or $\sin^2\theta_{23} \gsim 0.55$, the $\nu_\mu$ disappearance measurement has the leading power to determine $\sin^2\theta_{23}$ very accurately, apart from the disappearance ``octant'' degeneracy.

\item[Region II:] 
In a limited region inside $0.46 \lsim \sin^2\theta_{23} \lsim 0.55$, excluding a small region around $\sin^2\theta_{23} = 0.5$, the $\nu_e$ appearance measurement can constrain $\sin^2\theta_{23}$ better than the disappearance one. For particular combinations of the systematic errors in the appearance and disappearance channels this region may be absent, though (see for instance left panel in Fig.~\ref{fig:syst-dep-t2hk}).
The precise boundaries of this region with regions I and III (see below) depend very much on the size of the systematic errors, as one can see from Fig.~\ref{fig:syst-dep-t2hk}. 
\item[ Region III:] 
For values of $\sin^2\theta_{23}$ very close to maximal a third region appears, in which the disappearance measurement again supersedes the appearance measurement. As explained above, this is due to overlapping of the two clones when they are very close to maximal mixing. 

\end{description}

\noindent
Overall, one can see from Figs.~\ref{fig:syst-dep-t2hk} and~\ref{fig:syst-dep-lbne} that $\nu_e$ appearance and $\nu_\mu$ disappearance measurement cooperate to determine $\sin^2\theta_{23}$ very accurately, with a $1 \sigma$ uncertainty $0.02 - 0.03$, or $\sim 5\%$ level, which is comparable to the possible ultimate accuracy for $\sin^2\theta_{13}$ expected from reactor experiments. If $\theta_{23}$ is in Region I the error may be even smaller. We note, however, that Regions II and III are the ones to which the experimental results seem to be converging~\cite{Abe:2014ugx}.

\subsection{Appearance vs. disappearance channels in Neutrino Factory setting}
\label{sec:s23-NF}

\begin{figure}
\begin{center}
\includegraphics[scale=0.45]{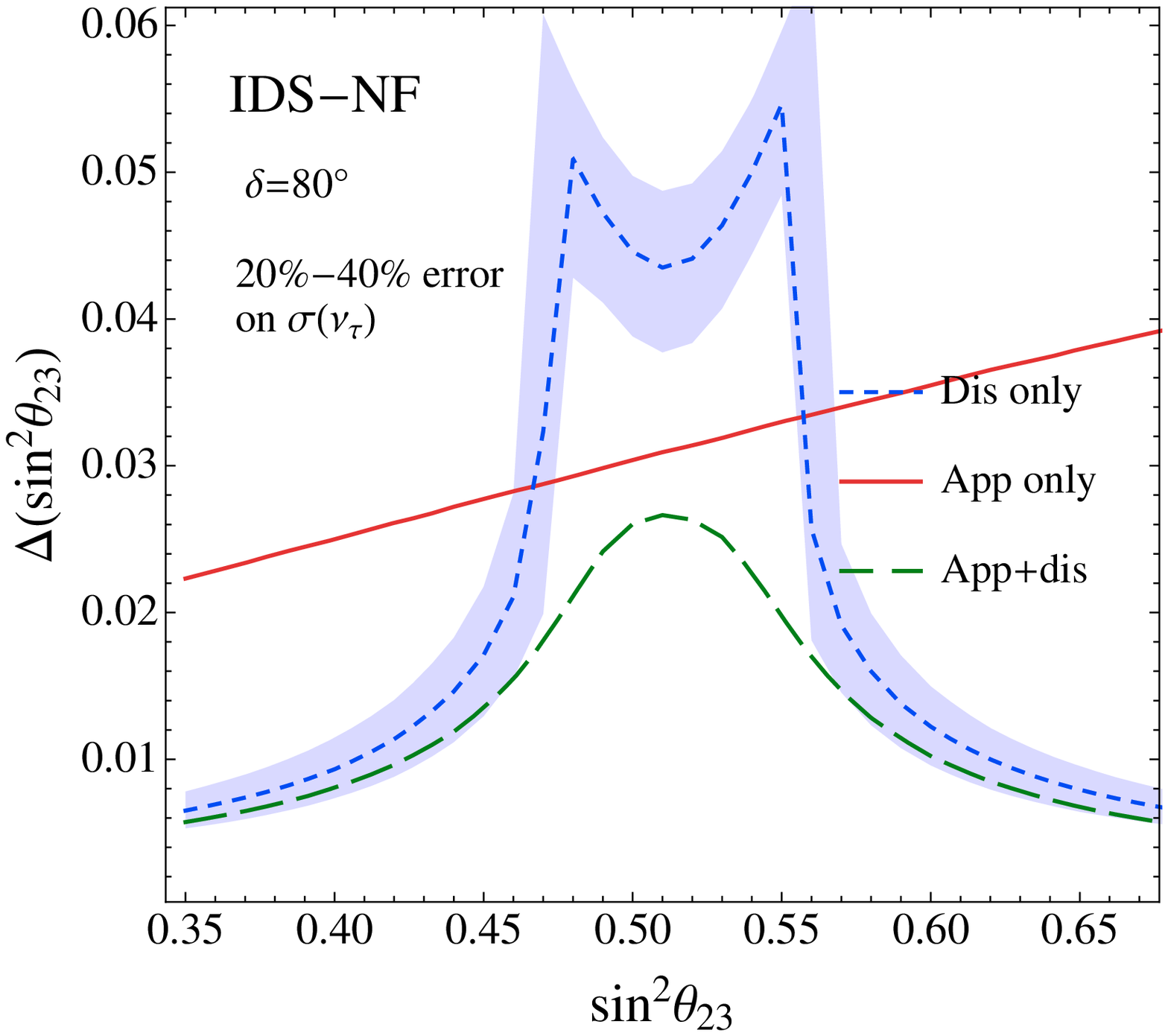}
\caption{Same as left panel in Fig.~\ref{fig:syst-dep-t2hk}, but for the IDS-NF setup. All lines correspond to the assumed uncertainty for this parameter in our default scenario (30\%), while the blue band shows the impact on the result for the disappearance channels if the systematic error on the $\nu_\tau$ cross section is varied between a 20\% and a 40\%. }
\label{fig:nufact}
\end{center}
\end{figure}

The relative importance of appearance and disappearance channels in determination of $\theta_{23}$ is quite different for the IDS-NF setup. As shown in Fig.~\ref{fig:nufact} Region III does not exist in this case, while Region II is quite wide, $0.44 \lsim \sin^2\theta_{23} \lsim 0.59$. Since the setting we consider for the neutrino factory is off VOM, the disappearance measurement is not as powerful as for facilities sitting at the first VOM like T2HK or LBNE. Also, note that the general features shown in Fig.~\ref{fig:nufact} are rather robust against variation of the systematic errors in the disappearance channel within a reasonable range, since for the NF this channel is mainly limited by being off-peak.

Finally, we have also examined the ESS$\nu$SB setting with a baseline of 540 km. Unfortunately, neither the disappearance nor the appearance measurement have sufficient statistics to determine $\sin^2\theta_{23}$ with a comparable accuracy to any of the other settings discussed above. For example, the appearance only measurement can reach only up to $\Delta (\sin^2\theta_{23}) \sim 0.07$ at $\sin^2\theta_{23} = 0.5$ for various input values of $\delta$. 

\subsection{Accuracy of measurements: $\sin^2\theta_{23}$ vs. $\sin \delta$}
\label{sec:23-delta}

Starting from simple analytical considerations, a simple expression relating the precision achievable for $\sth$ and $\sin\delta$ using {\em only} the appearance channel at the first VOM, was derived in Ref.~\cite{Minakata:2013eoa}:
\begin{equation}
\Delta (\sin^2\theta_{23}) \simeq \frac{1}{6} \Delta (\sin \delta).
\label{eq:error-relation}
\end{equation}

\begin{figure}
\begin{center}
\includegraphics[scale=0.5]{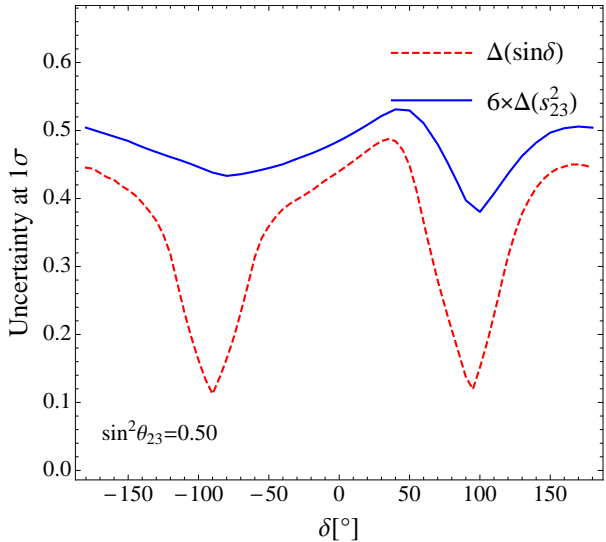} 
\caption{Comparison between the precision achievable for $\sth$ and the precision achievable for $\sin\delta$ for the T2HK setup. The solid line shows the error on $\sth$ multiplied by a factor of 6, while the dashed lines show the error on $\sin\delta$ as obtained directly from a simulation. The agreement between the two curves is noticeable in most of the $\delta$ parameter space, as predicted by Eq.~(\ref{eq:error-relation}), which was derived from simple analytical considerations in Ref.~\cite{Minakata:2013eoa} for facilities sitting at the first VOM. It should be noted that in this figure the \emph{full} size of the confidence interval is plotted in both cases, unlike for the rest of the figures in this paper where we show half of the size of the full confidence interval. }
\label{fig:error-relation}
\end{center}
\end{figure}

We have confirmed that this relation holds reasonably well when both observables are computed within the same experimental setup sitting near the VOM. The results are shown in Fig.~\ref{fig:error-relation} for the case of the T2HK setup. In this figure, the uncertainty on $\sin\delta$ is compared to the uncertainty on $\sth$ multiplied by a factor of 6. Results are shown as a function of the value of $\delta$ itself, for $\sth = 0.50$. 
As it can be seen from the figure, the agreement between the two curves is quite good, and they show a similar dependence with the value of $\delta$ itself, with the sole exception of the regions close to $\pm \pi/2$. 
%

The reason for the disagreement in these two regions can be partially explained by taking into account that the function $\sin\delta$ in these regions has an upper limit, while this is not the case for $\sth$ in the region under consideration (\ie, around maximal mixing). Therefore, one should expect the confidence interval in this region to be reduced by approximately a factor of 2 for $\sin\delta$. It is also related to the Jacobian involved in the measurement of $\sin\delta$, as partially discussed in Ref.~\cite{Coloma:2012wq}. 
The precision on $\sin\delta$ can be computed by doing a Taylor expansion by $\Delta(\delta)$, the uncertainty on $\delta$. To first order in the expansion, it gives $\Delta (\sin \delta) = \cos \delta \Delta (\delta)$, which implies that $\Delta (\sin \delta)$ should vanish at $\delta = \pm \frac{\pi}{2}$. 
When higher order terms are included in the Taylor expansion, however, a non-vanishing result is obtained, in agreement with the minima for the dashed curve in Fig.~\ref{fig:error-relation}. 

We have also examined whether the relation holds for different values of $\theta_{23}$. 
The qualitative features of the results are quite similar, but the difference between the solid blue and the dashed red curves in Fig.~\ref{fig:error-relation} becomes larger: it increases by approximately a factor of $\sim 3$ at $\delta=0$ when $\theta_{23}$ is varied from $\sth = 0.40$ to $\sth = 0.60$, while they are more similar near the dips ($\delta \sim \pm \pi/2$). 
%

\section{Determination of $\delta$}
\label{sec:delta}

\begin{figure}
\begin{center}
\begin{tabular}{cc}
  \includegraphics[scale=0.45]{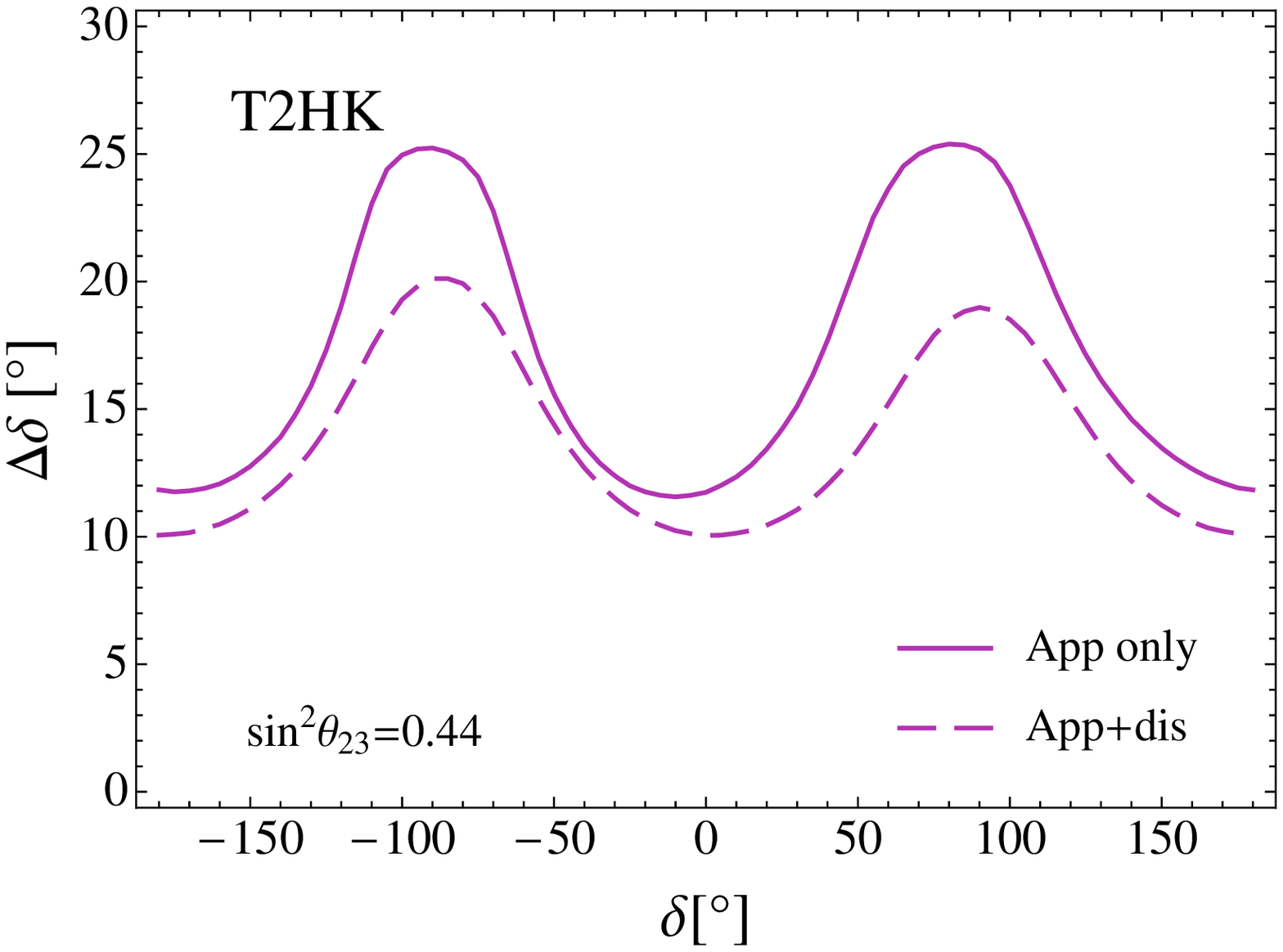} &
  \includegraphics[scale=0.45]{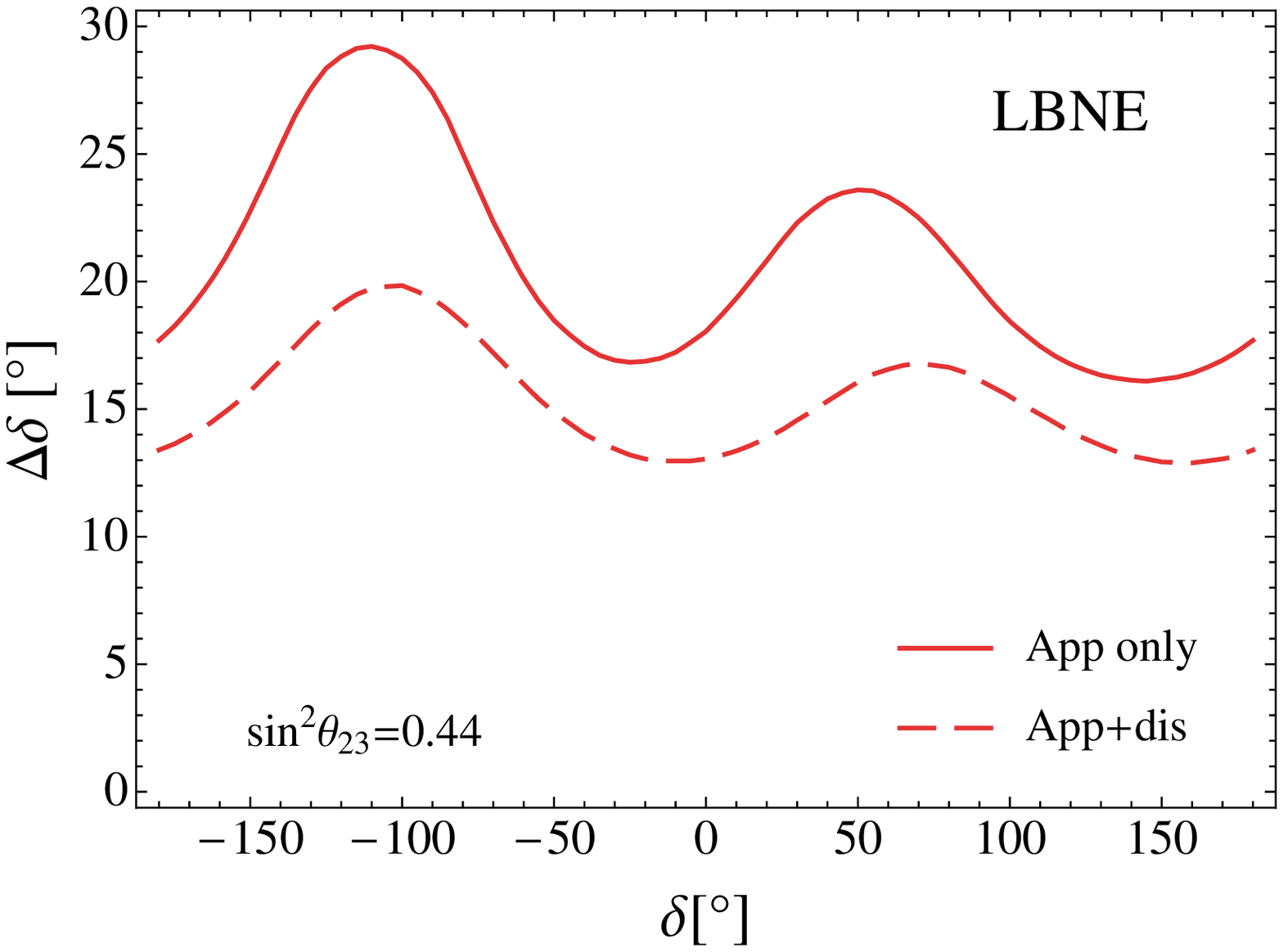} 
\end{tabular}
\caption{Precision on $\delta$ (at 1$\sigma$, for 1 d.o.f.) as a function of the value of $\delta$ itself, for T2HK in the left panel and for LBNE in the right panel. Solid lines show the results using the appearance channels only, while dashed lines show the results from the combination of appearance and disappearance data.  }
\label{fig:delta}
\end{center}
\end{figure}

Let us now explore what is the impact due to the combination of different channels on the determination of $\delta$. Figure~\ref{fig:delta} shows the expected precision for a measurement of $\delta$ for the T2HK (LBNE) experiment in the left (right) panel, as a function of the value of $\delta$ itself. Results are shown at 1$\sigma$ CL, for 1 d.o.f. As one can see, the addition of the disappearance channels is helping to get a better determination of $\delta$, specially in the regions around $\delta = \pm \pi/2$, as was already pointed out in Refs.~\cite{Donini:2005rn,Coloma:2012wq}. This effect comes mainly through a better determination of the squared mass splitting in the $\nu_\mu$ disappearance channels, which can be understood from the fact that $\delta$ and $\Delta m_{31}^2$ appear together in the appearance oscillation probability, see Eq.~\ref{Pemu-matter}. 

\begin{figure}
\begin{center}
\begin{tabular}{cc}
  \includegraphics[scale=0.45]{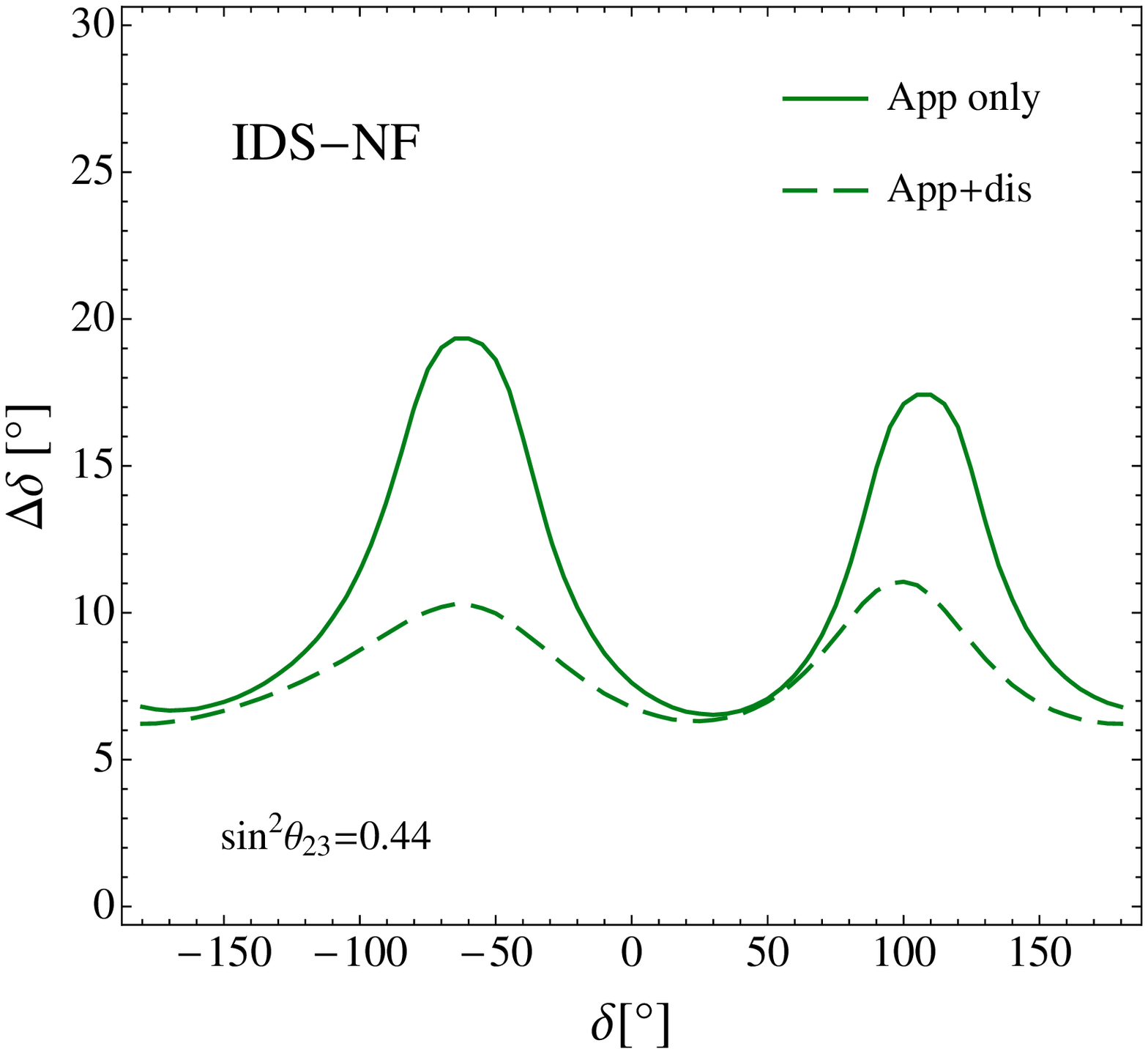}  &
  \includegraphics[scale=0.473]{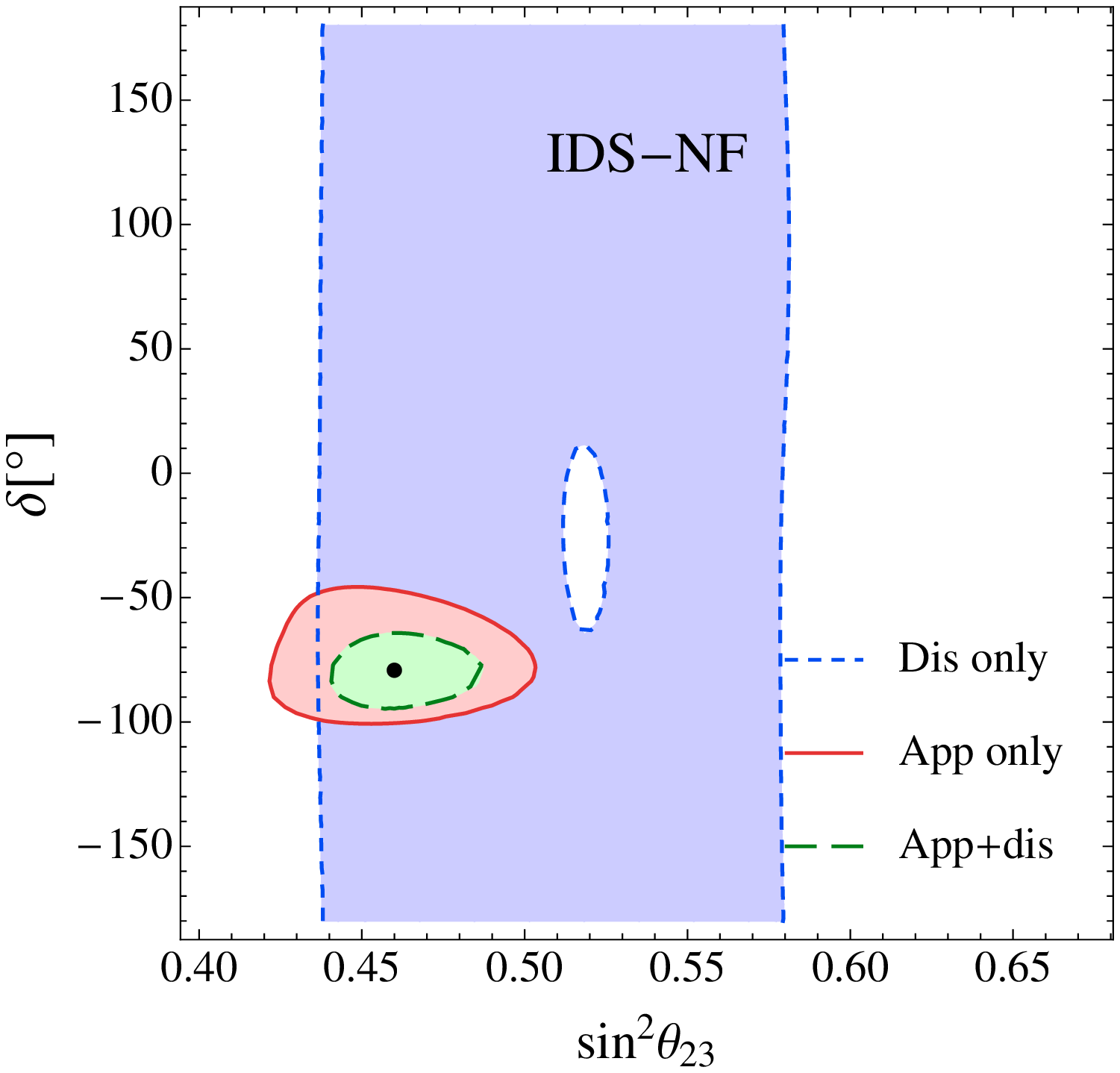} 
\end{tabular}
\caption{Left panel: expected precision on $\delta$ (at 1$\sigma$, for 1 d.o.f.)  as a function of the value of $\delta$ itself, for the IDS-NF setup.  Solid and dashed lines indicate the results obtained from the appearance results alone and from the combination between appearance and disappearance data, respectively. Right panel: confidence regions in the $\sth-\delta$ plane (at 1$\sigma$, for 2 d.o.f.) , for a particular set of true values and for different combinations of oscillation channels, as indicated in the legend. Results in both panels correspond to the IDS-NF setup as defined in Sec.~\ref{sec:exp}. It should be pointed out that the ``hole'' in the confidence region obtained for the disappearance channels vanishes just above the $1\sigma$ CL.
}
\label{fig:nufact-delta}
\end{center}
\end{figure}

The left panel in Fig.~\ref{fig:nufact-delta} shows similar results for the IDS-NF setup. The improvement on the determination of $\delta$ after the addition of disappearance channel data is remarkable for this setup. In the right panel in the same figure we show the confidence regions at 1$\sigma$ (2 d.o.f.) projected in the $\sth-\delta$ plane. Results are shown for the appearance channels alone (red region, solid line), disappearance channels alone (blue region, dotted line) and for the combination of appearance and disappearance channels (green region, dashed line). From this panel it can be clearly seen how the measurement of $\theta_{23}$ is coming mainly from the appearance channel for this setup, while the accurate determination of $\delta$ stems from the combination between appearance and disappearance data.

\begin{figure}
\begin{center}
\begin{tabular}{cc}
  \includegraphics[scale=0.45]{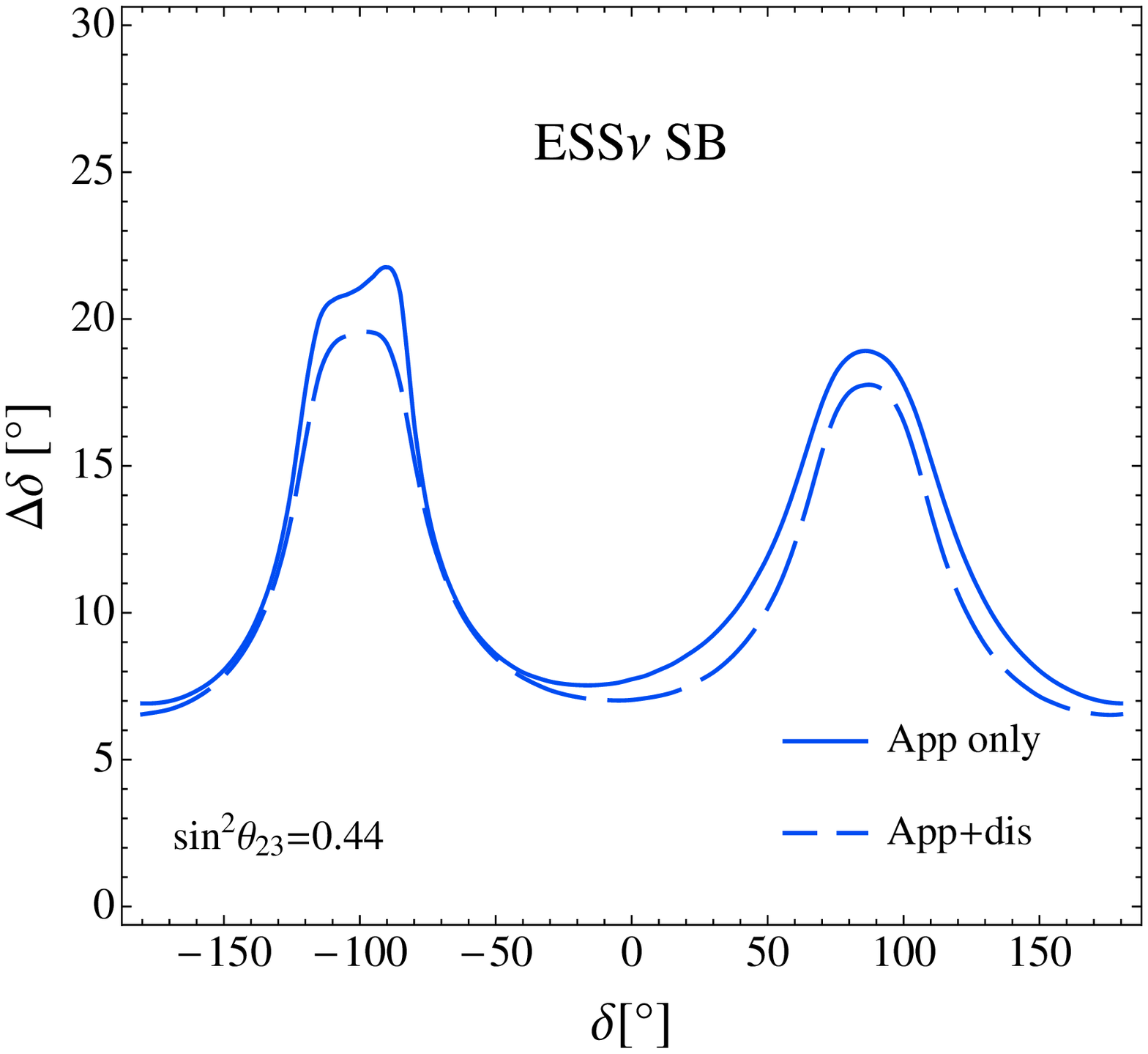}  &
  \includegraphics[scale=0.473]{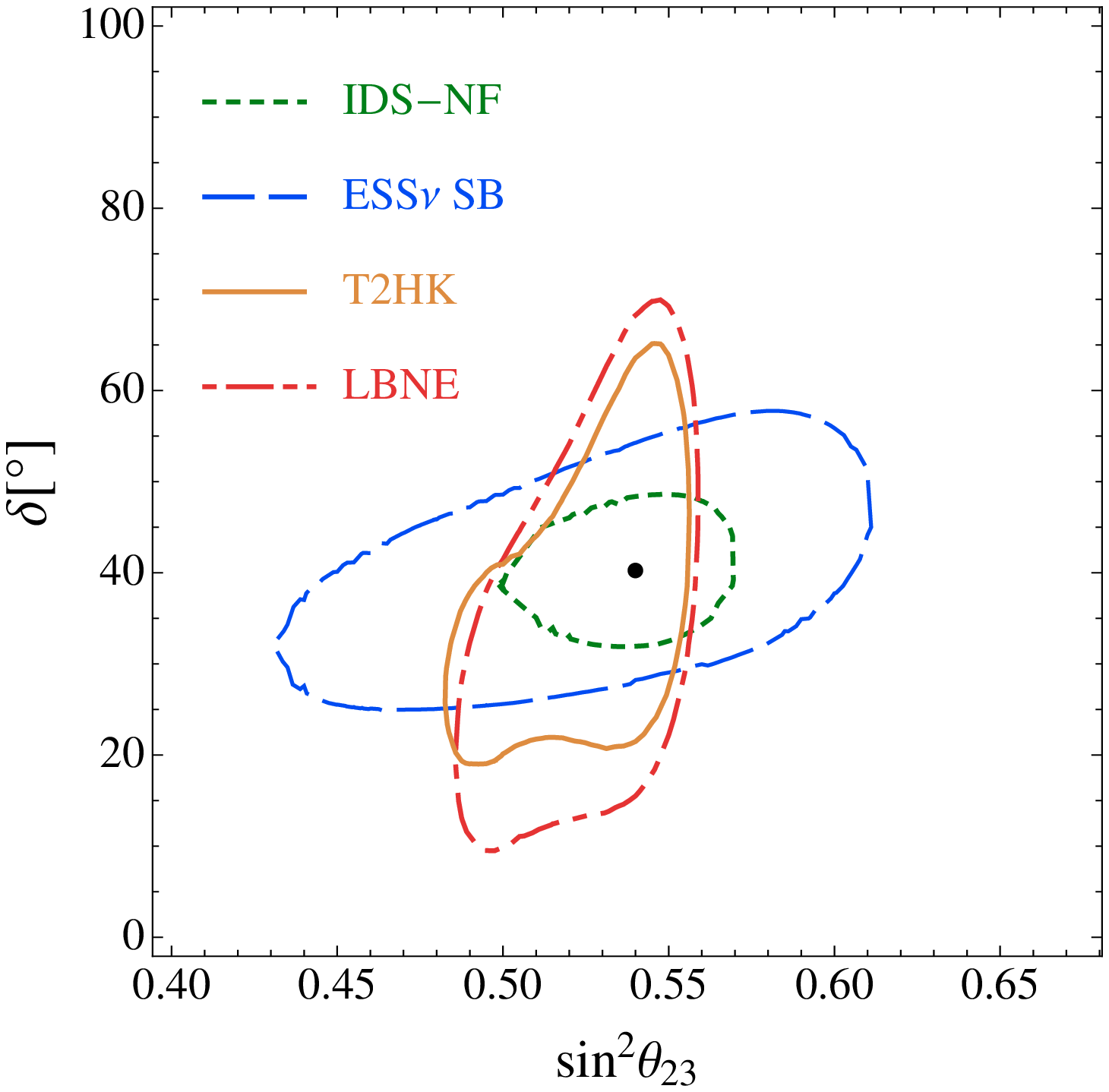} 
\end{tabular}
\caption{Left panel: expected precision on $\delta$ (at 1$\sigma$, for 1 d.o.f.) as a function of the value of $\delta$ itself for the ESS$\nu$SB setup. Solid and dashed lines indicate the results obtained from the appearance results alone and from the combination between appearance and disappearance data, respectively. Right panel: confidence regions in the $\sth-\delta$ plane (at $1\sigma$, for 2 d.o.f.), for a particular set of true values (indicated by the black dot) and for the four setups considered in this work. See Tabs.~\ref{tab:setups} and~\ref{tab:events} for a precise definition of the different setups and for the expected number of events in each oscillation channel, respectively. In this panel, all regions include data from both appearance and disappearance channels. }  
\label{fig:ess-delta}
\end{center}
\end{figure}

The last case under study in this section is the case of ESS$\nu$SB, for which the situation is very different from all the previous cases: 
The much smaller number of events at this facility would not allow to determine $\Delta m_{31}^2$ very precisely. Therefore, it is expected \emph{a priori} that for facilities operating at the second VOM the addition of disappearance data would be of little help in improving the accuracy of a measurement of $\delta$. This is confirmed by the results shown in the left panel of Fig.~\ref{fig:ess-delta}, the precision for $\delta$ obtained from appearance data alone (solid lines) and in combination with disappearance data (dashed lines). 
It is remarkable that, in spite of a factor of $\sim 50$ smaller number of appearance events in the ESS$\nu$SB than in IDS-NF (see Tab.~\ref{tab:events}) setups, the sensitivity to $\delta$ using only the appearance channels data is comparable with each other. It is the power of placing the detector at the second VOM where the dependence of the oscillation probability with $\delta$ is larger by a factor of three than that at the first VOM. It leads to an extremely good CP violation sensitivity as well as a very accurate determination of the value of $\delta$ and a reduced dependence on systematic errors, see Refs.~\cite{Marciano:2001tz,Ishitsuka:2005qi,Coloma:2011pg,Baussan:2013zcy}. 

Finally, in the right panel of Fig.~\ref{fig:ess-delta} we show the confidence regions in the $\sth-\delta$ plane at $1\sigma$ (2 d.o.f.) that would be obtained from the combination of appearance and disappearance data for the four facilities under study. The true values for $\sth$ and $\delta$ are indicated by the black dot. In all cases, our default values have been used for the systematic uncertainties, see App.~\ref{app:sim}. The first thing that can be noticed from this plot is the very different shape of the confidence regions for the different oscillation facilities. The ESS$\nu$SB allowed region (dashed blue line) is rather wide in the $\sth$ axis, while it gives extremely good sensitivity to $\delta$. The T2HK (solid yellow) and LBNE (dot-dashed red) regions are narrower along the $\sin^2\theta_{23}$ axis due to the disappearance constraint, but the measurement of $\delta$ would be less accurate.
From the shape of the confidence region it can also be observed that the disappearance degeneracy is affecting the determination of $\sin^2\theta_{23}$. Finally, the IDS-NF setup (dotted green) enables the most precise measurement of \emph{both} $\theta_{23}$ and $\delta$ due to a \emph{synergetic} combination of appearance and disappearance measurements.

\section{Summary and Conclusions}
\label{sec:conclusions}

Toward the completion of our understanding of the lepton flavor mixing, the right question to pose now is how to determine $\theta_{23}$ and the CP-violating phase $\delta$ at the same time and how their measurements are correlated. In this paper, we have addressed these questions. We did it in the context of four particular setups for proposed future facilities: T2HK, LBNE, IDS-NF, and ESS$\nu$SB. Throughout the paper we paid special attention to the interplay between the $\nu_\mu \rightarrow \nu_e$ and $\bar\nu_\mu \rightarrow \bar\nu_e$ appearance channels (or their T-conjugate channels for IDS-NF) and the $\nu_\mu$ and $\bar\nu_\mu$ disappearance channels. 

In the first part of this paper, we have analyzed structure of the parameter degeneracy which we would encounter in attempting a simultaneous measurement of $\delta$ and $\theta_{23}$. Despite the large number of previous works in the literature devoted to study degeneracies in neutrino oscillations, we found that the $\theta_{23} - \theta_{13} - \delta$ degeneracy has not been discussed in a general framework, which is mandatory if $\theta_{23}$ is close to maximal, as indicated by the recent measurements. 
We found that the general degeneracy boils down to the \emph{appearance} and \emph{disappearance} degeneracies. The former is a generalization of the $\theta_{13}$ and $\theta_{23}$ intrinsic degeneracies, whereas the latter is a generalization of what is usually called the $\theta_{23}$ octant degeneracy. Moreover, if $\theta_{23}$ is near maximal, the $\theta_{23}$ disappearance degeneracies join into a single region, aggravating the problem. We have discussed its characteristic features and illustrated some properties which are useful for its resolution in Sec.~\ref{sec:G-degeneracy}.

In the second part of this work, we have discussed the issue of appearance vs. disappearance measurement towards the determination of $\theta_{23}$ and $\delta$, and more importantly the interplay between them. Let us start with the measurement of $\theta_{23}$ by noting some of its generic features:

\begin{itemize}
\item
The precision on $\theta_{23}$ obtained from the $\nu_\mu$ disappearance channels alone generally shows a strong dependence on the size of the systematic errors. This is because the measurement is systematics dominated due to the high statistics. The error on $\sin^2 \theta_{23}$ always has a strong dependence on $\theta_{23}$ as well. It develops a ``bowler hat'' structure in region near the maximal $\theta_{23}$, which stems mainly from the merging clone effect, as we discussed in Sec.~\ref{sec:G-degeneracy}. We find that, if $\theta_{23}$ is far from maximal (in Region I as defined in Sec.~\ref{sec:relative}) the disappearance measurement always surpasses the appearance one in accuracy of determining $\sin^2 \theta_{23}$.

\item
On the other hand, the precision on $\theta_{23}$ obtained from the $\nu_{e}$ and $\bar{\nu}_{e}$ appearance measurement alone has a much weaker dependence on $\theta_{23}$ without suffering from the merging clone issue. The error is also less dependent on the size of systematic errors since these channels are mostly limited by statistics instead. 

\end{itemize}

We have studied the interplay between the appearance and disappearance oscillation channels at four particular setups: T2HK, LBNE, IDS-NF and ESS$\nu$SB. Their main features are summarized in Sec.~\ref{sec:exp}, while a more detailed description of the experimental setups can be found in App.~\ref{app:sim}. The expected number of events in each channel for all the facilities are summarized in Tab.~\ref{tab:events}. 
We observe the following:
\begin{itemize}

\item 
T2HK$/$LBNE:
Both of these facilities have values of $L/E$ very close to the first vacuum oscillation maximum (VOM). For both setups, the relative importance between the appearance and the disappearance channels for a precise determination of $\theta_{23}$ depends on the size of systematic errors. With reasonable estimates for the systematic uncertainties, we find that: (1) if $\sin^2 \theta_{23} \simeq 0.5$, the disappearance measurement gives slightly better sensitivity to $\sin^2 \theta_{23}$ than appearance; (2) for values of $\theta_{23}$ close to $\sin^2 \theta_{23} \simeq 0.49$ and $0.55$, typically, appearance measurement is more powerful than disappearance in most cases.
Despite their very different experimental setups, we find that both the expected accuracies as well as the features due to the interplay between the appearance and disappearance channels are very similar. 

\item 
IDS-NF: This setup would operate at an energy well above the first VOM. For values of $\sth$ such that $ 0.45 \lesssim \sth \lesssim 0.56$, the accuracy in the determination of $\theta_{23}$ comes mainly from the \emph{appearance} channels alone, while outside the mentioned interval the measurement is mainly driven by the disappearance channels. In particular, the precision for $\sin^2 \theta_{23}$ in the region very close to maximal mixing is worse than the one obtained through the appearance measurement by a factor of up to $3-4$. It has to do with the fact that the value of $L/E$ for the IDS-NF setup considered here turns out to be rather far from the oscillation maximum. Also, the disappearance measurement is largely affected by the systematic uncertainties on the $\tau$ backgrounds. We have found that the above result holds as long as the systematic errors associated to the $\nu_\tau$ charged-current cross section remain above the 20\%. 

\item 
ESS$\nu$SB: 
At the second VOM, the situation is quite different. In this case, the value of $L/E$ is tuned to maximize the impact of the interference term in the oscillation probability in order to obtain a better sensitivity to CP violation. The price to pay is that the sensitivity to $\theta_{23}$ is reduced, since it would mainly come from the precision measurement of the leading order term in the probability (for which a large number of events is needed).

\end{itemize}

We have studied the interplay of the different oscillation channels also on the determination of the CP phase $\delta$. Adding the disappearance channel generally improves the sensitivity to $\delta$, with the sole exception of ESS$\nu$SB. The improvement is always largest near $\delta \sim \pm \pi/2$. All the setups benefit from the addition of disappearance data mainly by a better determination of $\Delta m_{31}^2$, which allows cleaner discrimination of  effect of $\delta$ in the appearance channels. The size of the effect depends on various factors such as the number of events in the disappearance channels and the way the systematic errors are implemented, and hence it varies with the settings. 

For both the T2HK and LBNE setups considered in this work, we find that this effect is present to a similar degree, but more prominently for LBNE in particular outside the region $\delta \sim \pm \pi/2$. In the case of the IDS-NF setup, a great improvement is observed for the precision on $\delta$ as well, especially at around $\delta = \pm \pi/2$. 
It is evident from the right panel of Fig.~\ref{fig:nufact-delta} that the sensitivity to $\delta$ by the disappearance data itself is not impressive at all. Therefore, such a significant effect on sensitivity to $\delta$ must come from the synergy effect between the disappearance and appearance channels in the IDS-NF setup.

In the case of ESS$\nu$SB the situation is completely different. It is the unique case that essentially no improvement on the sensitivity to $\delta$ is achieved by adding the disappearance channel data. Yet, the precision in regions around $\delta \sim 0$ and $\pm \pi$ is remarkable, a high sensitivity that can be competed only by IDS-NF. On the other hand, the accuracy of $\delta$ determination at around $\delta \sim \pm \pi/2$ would be comparable to those of T2HK and LBNE. See Figs.~\ref{fig:delta},~\ref{fig:nufact-delta} and~\ref{fig:ess-delta}. 
It is worth mentioning that the comparable sensitivities to $\delta$ expected for the ESS$\nu$SB and IDS-NF setups using only appearance data are achieved with much smaller number of events (by a factor of $\sim 50$) at the former, indicating the power of the detector at the second VOM.

To conclude, we hope that the discussions given in this paper are useful to understand the physics behind the future precision measurement of $\theta_{23}$ and $\delta$, and that we will see some of the facilities described here realized in the near future.

\section*{Acknowledgments}

All authors would like to thank NORDITA and the organizers of the workshop ``NuNews: News in Neutrino Physics'', where part of this work was completed, for financial support and hospitality. P.C. would like to thank Enrique Fernandez-Martinez for providing the files needed to simulate the ESS$\nu$SB setup. P.C. and H.M. would like to thank the Fermilab Theory Group for hospitality during their visits. 
H.M. thanks Universidade de S\~ao Paulo for the great opportunity of
stay as Pesquisador Visitante Internacional. He is also partially supported by KAKENHI received through Tokyo Metropolitan University, Grant-in-Aid for Scientific Research No. 23540315, Japan Society for the Promotion of Science.
S.P.  acknowledges partial support from the  European Union FP7  ITN INVISIBLES (Marie Curie Actions, PITN- GA-2011- 289442). Fermilab is operated by the Fermi Research Alliance under contract no. DE-AC02-07CH11359 with the U.S. Department of Energy. Also this work has been partially supported by the U.S. Department of Energy under award number \protect{DE-SC0003915}.

\appendix

\section{Simulation details}
\label{app:sim}

The LBNE setup has been simulated following the Conceptual Design Report (CDR) from October 2012~\cite{CDR}, rescaling the beam power and detector size to the values listed in Tab.~\ref{tab:setups}. The neutrino fluxes correspond to 120~GeV protons. Data taking is set to a total of 10 years, equally split between $\nu$ and $\bar\nu$ modes. Migration matrices are used to account for the mis-reconstruction of Neutral Current (NC) events as Charged Current (CC) events at lower energies. The signal and the rest of the backgrounds are smeared in energy according to a Gaussian with $\sigma(E) = 0.15\times \sqrt{E}$ for electron neutrinos, and $\sigma(E) = 0.20\times \sqrt{E}$ for muon neutrinos. 

The T2HK setup has been simulated as in Ref.~\cite{Coloma:2012ji}. For this setup, both signal and backgrounds are reconstructed using the migration matrices from Ref.~\cite{Huber:2007em}. The beam power is set to 750~kW, and data taking is 10 years, divided between $\nu$ and $\bar\nu$ modes as indicated in Tab.~\ref{tab:setups}. Signal and background efficiencies have been adjusted to reproduce the number of events in the HK letter of intent (HK-LoI)~\cite{Abe:2011ts}.
\footnote{Note that in the HK-LoI the beam power is roughly a factor of 2 larger than the one used in this work. Nevertheless, the running time considered was a factor of two smaller and therefore the total number of events should be roughly the same.  } 
We have checked that our results for T2HK are roughly consistent with those in Refs.~\cite{Abe:2011ts,HK-sensitivity}. We have also checked that, for a reduced statistics, the results are roughly consistent with those reported by the T2K collaboration for the determination of $\sin^2\theta_{23}$, see Ref.~\cite{Abe:2014ugx}.

Regarding the ESS$\nu$SB, several possible configurations are currently under consideration. Here, we consider a setup in which the neutrino flux is produced from 2.5 GeV protons, and with a baseline accurately set to the second oscillation peak. This setup is simulated as in Refs.~\cite{Baussan:2012cw,Baussan:2013zcy}. The detector response is simulated using the migration matrices from Ref.~\cite{Agostino:2012fd}, which have been obtained for the MEMPHYS detector~\cite{deBellefon:2006vq}.

Finally, the IDS-NF setup considered here has been optimized for the large $\theta_{13}$ scenario, see Ref.~\cite{Agarwalla:2010hk}. This setup uses a MIND detector~\cite{Bayes:2012ex} placed at 2000~km from the source, see Tab.~\ref{tab:setups}. The MIND response is simulated using migration matrices for all signal and background contributions~\cite{nufactmatrices}. Backgrounds coming from $\tau$ decays~\cite{Indumathi:2009hg,Donini:2010xk} have also been included in the analysis.

Tab.~\ref{tab:events} shows the expected number of events for the setups described above. The two numbers in each column indicate the signal/background expected number of events, for a given setup and a given oscillation channel. Detector efficiencies have already been accounted for. In all cases, the same cross sections as in Ref.~\cite{Coloma:2012ji} have been used.
\begin{table}
\begin{center}
\renewcommand\tabcolsep{8pt}
\renewcommand\arraystretch{1.5}
\begin{tabular}{lc|cc|cc}
  & Energy range & $\nu$ app.  & $\bar\nu$ app. & $\nu$ dis. & $\bar\nu$ dis. \\ \hline
LBNE  & 0.5 - 8.0 GeV  & 1095/314 & 324/208 & 7340/82 & 3873/27  \\  
T2HK  & 0.4 - 1.2 GeV & 3984/1705 & 2161/1928 & 26237/716 & 19232/735  \\  
ESS$\nu$SB  & 0.1 - 1.0 GeV & 270/85 & 244/82 & 6198/113 & 4128/79 \\  
IDS-NF  & 0.1 - 9.0 GeV &  20241/476 & 5257/269 & 171133/7370 & 106077/3279  \\   \hline   
\end{tabular}
\caption{Number of events for the four setups considered in this work. The number of events for the signal/background component are given separately for each oscillation channel within a given setup, and detector efficiencies have already been accounted for. These event rates correspond to the following set of oscillation parameters: $\theta_{12}=32^\circ$, $\theta_{13}=9^\circ$, $\theta_{23}=45^\circ$, $\Delta m^2_{21}= 7.6 \times 10^{-5}\, \textrm{eV}^2$, $\Delta m^2_{31}= 2.45 \times 10^{-3} \, \textrm{eV}^2$ (normal ordering of neutrino masses). 
\label{tab:events} }
\end{center}
\end{table}

\section{The $\mathbf{\chi^2}$ and implementation of systematics uncertainties}

All results in Secs.~\ref{sec:relative} and~\ref{sec:delta} have been obtained using GLoBES~\cite{Huber:2004ka,Huber:2007ji}. The implementation of systematics has been done using a modified version of GLoBES, as in Ref.~\cite{Coloma:2012ji}. The $\chi^2$ and systematic uncertainties are implemented as follows. For each energy bin $i$ a contribution to the $\chi^2$ is computed as:
\begin{equation}
\chi^2_{i} (\theta;\xi) =  2\bigg(T_{i}(\theta;\xi)-O_{i}+O_{i} \ln \frac{O_{i}}{T_{i}(\theta;\xi))} \bigg) \,,\\
\end{equation}
where $O_{i}$ stands for the observed (true) event rates, and 
\begin{eqnarray}
T_{i}(\theta;\xi) & = & \left[1 + \xi_{\phi,i}\right] s_{\nu,i}(\theta) + 
\left[1 + \xi_{bg,\nu,i}  \right]b_{\nu,i} \nonumber \\ 
& + & \left[1 + \xi_{\phi,i} + \xi_{\bar\nu/\nu}\right] s_{\bar\nu,i}(\theta) + 
\left[1 + \xi_{bg,\bar\nu,i}  \right]b_{\bar\nu,i}
\label{eq:T}
\end{eqnarray}
corresponds to the true (fitted) event rates observed in the $i$-th energy bin for a given oscillation channel. Here, $\theta$ indicates the dependence on the test values for the oscillation parameters. It should be noted that $O_{i}$ depends only on the true values assumed for the oscillation parameters, while $T_{i}$ depends on the pair of values we are testing as well as on the nuisance parameters. $ \xi_{\phi,i}$ stands for the nuisance parameter associated to a combination of flux and cross section uncertainties for the signal. We take this uncertainty to be correlated between neutrinos and antineutrinos within the same oscillation channel. $\xi_{\bar\nu/\nu}$ is a relative normalization uncertainty included \emph{only} in the antineutrino channels, which accounts for the difference between neutrino and antineutrino cross section uncertainties. Finally, $\xi_{bg,\nu,i}$ and $\xi_{bg,\bar\nu,i}$ correspond to the background normalization uncertainties in the neutrino and antineutrino channels. Note that the normalization uncertainty $\xi_{\bar\nu/\nu}$ is correlated among all energy bins; however, the rest of the nuisance parameters are allowed to vary independently for each bin during marginalization to account for shape uncertainties.

The final $\chi^2$ needs to be minimized over the nuisance parameters. It reads:
\begin{equation}
\chi^2 (\theta) = \textrm{min}_{\xi}\left\{\sum_{i}\chi^2_{i}(\theta;\xi)+\left(\frac{\xi_{\phi,i}}{\sigma_\phi}\right)^2+\left(\frac{\xi_{\bar\nu/\nu}}{\sigma_{\bar\nu/\nu}}\right)^2 + \left(\frac{\xi_{\nu,bg,i}}{\sigma_{\nu,bg}}\right)^2 + \left(\frac{\xi_{\bar\nu,bg,i}}{\sigma_{\bar\nu,bg}}\right)^2 \right\},\\
\label{eq:chi2}
\end{equation}
where the three last terms are the pull-terms (penalty terms) associated to the nuisance parameters, and the $\sigma_k$ are the prior uncertainties assumed for each systematic error $\xi_k$. Unless otherwise stated, for conventional neutrino beams we set the priors on the systematic uncertainties to the following values:
\bea
 \sigma_{\phi} = 5\% ; & \sigma_{\bar\nu/\nu} = 10\% ; & \sigma_{bg} = 10\% \, .
\eea
In some cases we will show how the results change when $\sigma_{\phi}$ is increased to 15\% (see Figs.~\ref{fig:syst-dep-t2hk} and \ref{fig:syst-dep-lbne}). We have checked that this is the prior uncertainty which generally has the larger impact on the results of our analysis for the LBNE and T2HK setups.

The IDS-NF, on the other hand, is less affected by systematic errors: any beam related uncertainties will be small since the flux can be computed analytically. Moreover, the availability of both electron and muon neutrino flavors at the near detector would allow to determine their cross sections very precisely. We have chosen to use the same systematics implementation (Eqs.~\ref{eq:T} and \ref{eq:chi2}) for this facility as for conventional beam experiments in order to ease the comparison between different facilities. However, in this case we assume that the near detector will generally do a better job cancelling systematic uncertainties and therefore use the following priors:
\begin{equation}
 \sigma_{\phi} = 3\% ; \qquad \sigma_{\bar\nu/\nu} = 5\% ; \qquad \sigma_{bg} = 10\%; \qquad \sigma_{\tau} = 30\% \, .
\end{equation}
Here, $\sigma_{\tau}$ refers in particular to the prior uncertainty associated to the $\nu_\tau$ interaction cross section (which affects the backgrounds coming from $\tau$ contamination only~\cite{Indumathi:2009hg,Donini:2010xk}), while $\sigma_{bg}$ is used for all the other background contributions. We have checked that the $\sigma_{\tau}$ prior uncertainty has the largest impact on the results for the IDS-NF setup used in this work. Its impact on the results for $\sth$ is shown in Fig.~\ref{fig:nufact}.

Finally, marginalization is also performed over the oscillation paramteters and the matter density. Unless otherwise stated, the following true values are assumed for the solar mixing parameters, $\theta_{13}$ and the atmospheric mass splitting:
\begin{equation}
\begin{array}{rclcrcl}
\theta_{12} & = & 33.2^\circ  & \phantom{lala} & \Delta m_{21}^2 & = & 7.50\times 10^{-5} \, \textrm{eV}^2 \, , \nonumber \\
\theta_{13} & = & 9.2^\circ & \phantom{lala} & \Delta m_{31}^2  & = & 2.4\times 10^{-3} \, \textrm{eV}^2 \, (\textrm{NH}) \, , \nonumber 
\end{array}
\end{equation}
while the assumed true values of $\theta_{23}$ and $\delta$ will be specified in each case. NH stands for normal hierarchy, \ie, $m_3>m_1$. For each of the parameters which are marginalized over, a penalty term is added to the $\chi^2$ in Eq.~\ref{eq:chi2} in the same way as it was done for the systematic uncertainties, and the global minimum is searched for. 

Marginalization is always performed over the oscillation parameters not shown in each plot, using Gaussian priors with the following $1\sigma$ errors: 3\% for the solar oscillation parameters; a 4\% for the atmospheric mass splitting; 0.005 for $\sin^22\theta_{13}$ and 0.08 for $\sin^22\theta_{23}$. The values chosen for the solar and atmospheric mass splitting are in agreement with the current $1\sigma$ errors from global fits, see for instance Ref.~\cite{Tortola:2012te,GonzalezGarcia:2012sz,Capozzi:2013csa,nufit}. For $\sin^22\theta_{13}$ we have used the precision expected at the end of the running of Daya Bay, assuming it is limited by their systematic error~\cite{Dwyer:2013wqa}. For $\sin^22\theta_{23}$, on the other hand, we use a value which lies roughly in between the current precision achieved at T2K and MINOS and the T2K systematic uncertainty for this parameter, see Refs.~\cite{Abe:2012gx,Adamson:2011ig}. Unless otherwise stated, $\delta$ is left completely free during marginalization (\ie, no prior is assumed for this parameter). Finally, the value of the matter density is set according to the PREM profile~\cite{prem,Dziewonski:1981xy},and a 2\% prior uncertainty is assumed for this parameter.


\end{document}